\documentclass[twocolumn]{aa}
\usepackage{natbib}
\bibpunct{(}{)}{;}{a}{}{,} % to follow the A&A style
\usepackage{graphicx}
\usepackage{txfonts}
\usepackage{xspace}
\usepackage{supertabular}
\usepackage{longtable}
\usepackage[version=3]{mhchem}
\usepackage[OT2,OT1]{fontenc}
\usepackage{mathtools}
\usepackage{hyperref}
\usepackage{siunitx}
\usepackage{xcolor}
\newcommand\cyr
{
	\renewcommand\rmdefault{wncyr}
	\renewcommand\sfdefault{wncyss}
	\renewcommand\encodingdefault{OT2}
	\normalfont
	\selectfont
}
\DeclareTextFontCommand{\textcyr}{\cyr}

\newcommand{\kms}{km~s$^{-1}$\xspace}
\newcommand{\um}{$\mu$m\xspace}

\begin{document}

   \title{Hot Exoplanet Atmospheres Resolved with Transit Spectroscopy (HEARTS)}
    \subtitle{IV. A spectral inventory of atoms and molecules in the high-resolution transmission spectrum of WASP-121 b}

   \author{H.~J. Hoeijmakers\inst{1,2,3}
	 	  \and
	 	  J.~V. Seidel\inst{1}
          \and
          L. Pino\inst{4,5}
          \and
          D. Kitzmann\inst{2}
          \and
          J.~P. Sindel\inst{6,7,8}
          \and
          D. Ehrenreich\inst{1}
          \and
          A.V. Oza\inst{9}
          \and
		  V. Bourrier\inst{1}
		  \and
          R. Allart\inst{1}
          \and
		  A. Gebek\inst{10}
		  \and
          C. Lovis\inst{1}
          \and
          S.N. Yurchenko\inst{11}
          \and
          N. Astudillo-Defru\inst{12}
          \and
          D. Bayliss\inst{13}
          \and
          H. Cegla\inst{1}
          \and
          B. Lavie\inst{1}
          \and
          M. Lendl\inst{1}
          \and
          C. Melo\inst{14}
          \and
          F. Murgas\inst{15,16}
          \and
          V. Nascimbeni\inst{17}
          \and
          F. Pepe\inst{1}
          \and
          D. S\'egransan\inst{1}
          \and
          S. Udry\inst{1}
          \and
          A. Wyttenbach\inst{18}
          \and
          Kevin Heng\inst{2,13}
   }

   \institute{Observatoire de Gen\`eve, Universit\'e de Gen\`eve,
              51 Chemin des Maillettes, 1290 Sauverny, Switzerland
              \and
             Center for Space and Habitability, Universit\"at Bern, Gesellschaftsstrasse 6, 3012 Bern, Switzerland\\
             \email{jens.hoeijmakers@space.unibe.ch}
             \and
             Lund Observatory, Department of Astronomy and Theoretical Physics, Lunds Universitet, Solvegatan 9, 222 24 Lund, Sweden
             \and
             Anton Pannekoek Institute of Astronomy, Universiteit van Amsterdam, Science Park 904, 1098 XH Amsterdam, The Netherlands
             \and
             INAF-Osservatorio Astrofisico di Arcetri Largo Enrico Fermi 5, I-50125 Firenze, Italy
             \and
             Institute of Astronomy, Katholieke Universiteit Leuven,
             Celestijnenlaan 200D
             3001 Leuven, Belgium
             \and
            Centre for Exoplanet Science, University of St Andrews, North Haugh, St. Andrews, KY169SS, United Kingdom
            \and
            SUPA, School of Physics \& Astronomy, University of St. Andrews, North Haugh, St Andrews, KY169SS, United Kingdom
            \and
            Physikalisches Institut, Universit\"at Bern, Gesellschaftsstrasse 6, 3012 Bern, Switzerland
            \and
           Department of Physics, ETH Z\"urich, Wolfgang-Pauli-Strasse 27, 8093 Z\"urich, Switzerland
            \and
             Department of Physics \& Astronomy, University College London, Gower Street, London WC1E 6BT, United Kingdom
             \and
             Department of Mathematics and applied Physics, Universidad Cat\'olica de la Sant\'isima Concepci\'on, Alonso de Rivera 2850, Concepci\'on, Chile
             \and
             Department of Physics,
             University of Warwick,
             Coventry CV4 7AL,
             United Kingdom
             \and
             European Southern Observatory, Alonso de Cordova 3107, Vitacura, Regin Metropolitana, Chile
             \and
             Instituto de Astrofísica de Canarias (IAC), 38205 La Laguna, Tenerife, Spain
             \and
             Departament of Astrophysics, Universidad de La Laguna (ULL), 38206, La Laguna, Tenerife, Spain
             \and
            Department of Physics and Astronomy, Universit\`a degli Studi di Padova, Vicolo dell'Osservatorio 3, I-35122 Padova, Italy
             \and
             Universit\'e Grenoble Alpes, CNRS, IPAG, 38000 Grenoble, France}

   \date{Received May 7, 2020; accepted June 19, 2020}

  \abstract
   {WASP-121 b is a hot Jupiter that was recently found to possess rich emission (day side) and transmission (limb) spectra, suggestive of the presence of a multitude of chemical species in the atmosphere.}
   {We survey the transmission spectrum of WASP-121 b for line-absorption by metals and molecules at high spectral resolution, and elaborate on existing interpretations of the optical transmission spectrum observed with HST/STIS and WFC3.}
  % methods heading (mandatory
   {We apply the cross-correlation technique and direct differential spectroscopy to search for sodium and other neutral and ionised atoms, TiO, VO and SH in high-resolution transit spectra obtained with the HARPS spectrograph. We inject models assuming chemical and hydrostatic equilibrium with varying temperature and composition to enable model comparison, and employ two bootstrap methods to test the robustness of our detections.}
   {We detect neutral \ion{Mg}{}, \ion{Na}{}, \ion{Ca}{}, \ion{Cr}{}, \ion{Fe}{}, \ion{Ni}{} and \ion{V}{}{}, which we predict exists in equilibrium with a significant quantity of VO, supporting earlier observations by HST/WFC3. Non-detections of \ion{Ti}{} and TiO support the hypothesis that \ion{Ti}{} is depleted via a cold-trap mechanism, as has been proposed in the literature. Atomic line depths are under-predicted by hydrostatic models by a factor of 1.5 to 8, confirming recent findings that the atmosphere is extended. We predict the existence of significant concentrations of gas-phase TiO$_2$, VO$_2$ and TiS, which could be important absorbers at optical and near-IR wavelengths in hot Jupiter atmospheres, but for which accurate line-list data is currently not available. We find no evidence for absorption by SH, and find that inflated atomic lines can plausibly explain the slope of the transmission spectrum observed in the near-UV with HST/STIS. The \ion{Na}{I} D lines are significantly broadened (FWHM $\sim$ 50 km~s$^{-1}$\xspace to 70 km~s$^{-1}$\xspace) and show a difference in their respective depths of $\sim15$ scale heights, which is not expected from isothermal hydrostatic theory. If this asymmetry is of astrophysical origin, it may indicate that \ion{Na}{I} forms an optically thin envelope reminiscent of the \ion{Na}{I} cloud surrounding Jupiter, or that it is hydrodynamically outflowing.}
    {}
   \maketitle

\section{Introduction}
Gas giant exoplanets that orbit sufficiently closely to their host star exhibit elevated equilibrium temperatures that significantly alter their atmospheric structure when compared to cooler hot Jupiters. The atmospheres of these so-called ultra-hot Jupiters (UHJs) are subject to thermal dissociation of molecules and partial thermal ionization of atomic species. At temperatures over $2,500 - 3,000~$K, atomic hydrogen becomes the dominant atmospheric constituent \citep{Lothringer2018,Kitzmann2018,Parmentier2018}. Dissociation of hydrogen reduces the mean molecular weight (increasing the pressure scale height), and electrons liberated by the thermal ionization of metals (mostly alkalis) combine with atomic hydrogen to form H$^-$, which is a strong source of continuum opacity \citep{Arcangeli2018}. Indeed, the dominance of atomic hydrogen and the resulting importance of H$^-$ in shaping the radiative properties of the atmosphere may be a defining characteristic of UHJs.\\

\noindent UHJs are prime targets for observations that target the transmission spectrum of the atmosphere during transit events. The atmospheric scale height is large due to the high temperature and low mean molecular weight, while the short orbital period allows transits to be observed frequently. From a theoretical point of view, the atmospheres of the hottest UHJs reside in an interesting chemical regime. An absence of complex molecular chemistry precludes the formation of aerosol particles and opaque cloud decks, that have mostly defied chemical characterization when encountered in the atmospheres of cooler hot Jupiters. Combined with the fast rates of chemical reactions prevalent at high temperatures, the absence of complex molecular chemistry simplifies the theoretical interpretation of transmission/emission spectroscopy of these objects, making them especially amenable for detailed chemical characterization.\\

\noindent The UHJ KELT-9 b is the hottest planet in this class currently known \citep{Gaudi2017}. With an equilibrium temperature of over 4,000 K and a night-side temperature greater than 3,000 K \citep{Wong2019}, its atmosphere is expected to be nearly fully dissociated and in chemical equilibrium \citep{Lothringer2018,Kitzmann2018,Parmentier2018}, thereby epitomising the defining characteristics of UHJs. The atmospheres of cooler UHJs span a transition: from being dominated by molecular hydrogen at lower temperatures to being mostly dissociated and partially ionized \citep{Lothringer2018}. Due to the large day-to-night side temperature contrast, the atmosphere of any single UHJ may span multiple regimes itself, where the atmosphere on the strongly heated day side may be thermally dissociated whereas the cooler night side is instead governed by condensation processes and molecular chemistry \citep{Parmentier2018,Ehrenreich2020}.\\

\noindent WASP-121 b orbits a relatively bright (V=10.5) F6V star in a 1.27 day period orbit \citep{Delrez2016}.
The equilibrium temperature was estimated at $2358 \pm 52$ K \citep{Delrez2016}, prompting \citet{Evans2016} to search for the presence of TiO/VO and an atmospheric inversion layer \citep{Hubeny2003,Burrows2007}, using the WFC3 instrument aboard the Hubble Space Telescope (HST). Transit observations between 1.12 - 1.64$~\mu$m and optical spectro-photometry indicated clear absorption by water that is typically observed in hot Jupiter transmission spectra \citep{Sing2016}, as well as additional opacity near 1.2$~\mu$m tentatively ascribed to a combination of TiO/VO and FeH \citep{Evans2016}. Secondary-eclipse observations using the G141 grism of WFC3 (1.12$\mu$m - 1.64 $\mu$m) revealed emission by water on the day side hemisphere of the planet, indicating the presence of a temperature inversion \citep{Evans2017}. An additional source of emission near 1.2 $\mu$m was tentatively attributed to vanadium-oxide (VO).\\

\noindent To confirm the importance of VO, water and an inversion layer, \citet{Evans2018,Evans2019} and \citet{Evans2020} obtained repeated HST observations of the transmission spectrum and the secondary eclipse using the STIS and WFC 3 instruments. The optical transmission spectrum displays rich variation, with multiple features consistent with VO absorption that \citet{Evans2018} could reproduce by assuming an isothermal T-P profile at 1,500 K and a metallicity equivalent to $10\times$ to $30\times$ solar. Absorption bands of TiO appeared to be muted in the transmission spectrum, which was explained by \citet{Evans2018} as evidence of condensation of Ti-bearing species, which commences at higher temperatures than condensation of V-bearing species, producing e.g. calcium titanates \citep{Lodders2002} while VO remains in the gas phase. \citet{Evans2019} observed the the day side emission spectrum with the G102 grism of WFC3 (0.8$~\mu$m to 1.1$~\mu$m), augmenting their earlier observations with the G141 grism. The G102 spectrum does not show the VO bands expected to be present there, and this led \citet{Evans2019} to question the interpretation that the 1.2$~\mu$m feature is caused by VO emission. The secondary eclipse was observed at 2~\um \citep{Kovacs2019} and at optical wavelengths with the TESS instrument. These were analyzed together with the preceding Hubble, Spitzer and ground-based observations to yield tighter constraints on the atmospheric structure, composition and overall system parameters \citep{Bourrier2019,Daylan2019}. These studies found that the hottest point on the day side exceeds a temperature of 3,000 K, that the atmosphere is inverted on the day side, and a metallicity that is consistent with solar \citep{Bourrier2019} or slightly elevated \citep{Daylan2019}. Although the chemical retrievals follow different strategies (equilibrium versus free-chemistry), both indicate that a depletion of TiO relative to VO is needed to explain the observed emission spectrum, supporting the earlier findings by \citet{Evans2019}. Recently, \citet{Evans2020} obtained new secondary-eclipse observations using the G141 grism of WFC3. Although confirming the presence of emission by H$_2$O, a joint analysis with their previous WFC3 observations did not reproduce the emission feature at 1.2 \um, prompting the authors to entirely discard their previous interpretation of emission caused by VO.\\

\noindent At shorter wavelengths covered by the STIS instrument, \citet{Evans2018} observed a steep increase of the transit radius, which they propose to explain as being caused by the NUV absorption bands of an unknown absorber, and explore the possibility of the SH molecule. SH has been proposed to be a significant by-product of photo-dissociation of H$_2$S \citep{Zahnle2009,Zahnle2016}. Despite being a highly reactive radical, its abundance could exceed 1 ppm at the millibar level where transit transmission spectroscopy is sensitive. \citet{Evans2018} are able to reproduce the NUV slope with atmospheric models spanning temperatures of 1,500 to 2,000 K and SH abundances of 20 to 100 ppm, but note that an unambiguous identification of the SH molecule is beyond the potency of these low-resolution spectra. The transmission spectrum was observed at UV wavelengths using SWIFT/UVOS between 200 and 270 nm, yielding a tentative excess in the photometric transit depth that evidences strongly absorbing metal ions at high altitudes \citep{Salz2019}. Two transits observed using the high-resolution E230M echelle grating of HST/STIS between 228 and 307 nm revealed strong absorption lines by \ion{Fe}{II} and \ion{Mg}{II} \citep{Sing2019}. These absorption line depths are significantly greater than the transit radius of the planet's Roche lobe, indicating that these heavy elements may be part of a hydrodynamic outflow.\\

\noindent The planet atmosphere has also been observed at high spectral resolution at optical wavelengths, with the HARPS \citep{Bourrier2020,Cabot2020} and UVES \citep{Gibson2020,Merritt2020} spectrographs. These observations yielded confident detections of atomic metals, including \ion{Fe}{I} and \ion{Na}{I}, as well as H$-\alpha$ absorption - the latter of which is further evidence for the existence of an extended outflowing envelope. Given that \ion{Fe}{I} is notably rich in absorption lines at blue-optical and NUV wavelengths, these observations suggest that it could be responsible for the heating required to generate the observed temperature inversion in the upper atmosphere \citep{Gibson2020,Pino2020}, and may also explain the observed slope towards NUV wavelengths \citep{Lothringer2020}.\\

\noindent \citet{Merritt2020} further investigated the UVES spectra published by \citet{Gibson2020} in search for TiO and VO absorption. These authors report non-detections of both molecules, with an upper limit on the TiO abundance of $[\text{TiO}] \lesssim -9.3$, consistent with the retrieval of \citet{Evans2018} and the interpretation that TiO is condensed out of the gas phase. \citet{Merritt2020} also establish an upper limit on the VO abundance, but note that existing line-list are likely not accurate enough for application with high-resolution spectroscopy.\\

\noindent High-resolution spectrographs with $R=\frac{\lambda}{\Delta \lambda} = \frac{c}{\Delta v} \sim 10^5$ like HARPS and UVES are typically able to measure the centroid velocities of atmospheric signatures with sensitivities on the order of 1 \kms, making such observations sensitive to global atmospheric dynamics \citep{Snellen2010,Brogi2016,Flowers2019}. \citet{Bourrier2020,Cabot2020} and \citet{Gibson2020} independently report blueshifts of $-5.2 \pm 0.5$, $-3_{-1}^{+3}$ and $-4.4 \pm 0.6$ \kms, indicating the presence of a wind that carries these atoms from the day side to the night side at high altitudes. \\

\noindent In this paper we present an analysis of the high-resolution transmission spectrum of WASP-121 b, using the three transit events observed with the HARPS spectrograph first described by \citet{Bourrier2020}. We apply a classical differential analysis targeting the strong sodium (\ion{Na}{I}) doublet following the strategy of \cite{Wy15,Se19} as well as the cross-correlation technique \citep{Snellen2010} to search for additional atomic and ionized metals, following the analyses of \citet{Hoeijmakers2018} and \citet{Hoeijmakers2019}. Section 2 describes the observations and details the transmission spectroscopy and cross-correlation analyses. Section 3 presents the detected species and discusses their implications for the chemical composition and structure of the atmosphere of WASP-121 b. Section 4 enumerates the conclusions with reference to the most important figures and tables. Appendices \ref{app:analysis_steps}, \ref{app:bootstrap} and \ref{app:results} provide supporting Figures of the cross-correlation procedure, all obtained cross-correlation functions and a detailed description of our bootstrap methods used to assess detection robustness.

\section{Observations and analysis process}\label{sec:obs}
\subsection{HARPS observations}\label{sec:harps}
We obtained spectra during three transits of the hot gas giant planet WASP-121b around its host star WASP-121 (spectral type F6, V=$10.4$), with the HARPS instrument at the ESO 3.6 m telescope in La Silla Observatory, Chile \citep{Ma03}. The observations were performed on the 31st Dec 2017, the 9th Jan 2018 and 14th Jan 2018 as part of the HEARTS survey (ESO programme: 100.C-0750; PI: Ehrenreich). A log of the observations is provided in Table \ref{table:nightoverview}. Flat field and wavelength calibration frames are obtained during the daily afternoon calibration. All science observations are performed with fibre A on the target and fibre B on the sky.

 \begin{table*}
\caption{Log of observations.}
\label{table:nightoverview}
\centering
\begin{tabular}{c c c c c c c }
\hline
\hline
&Date   &$\#$Spectra \tablefootmark{a}  &Exp. Time [s]  &Airmass        \tablefootmark{b}&Seeing        &SNR order 56 \tablefootmark{c}      \\
\hline
Night 1&2017-12-31&35 (16/19)&650&1.4-1.05-1.3&0.7-1.2&27 -  63\\
Night 2&2018-01-09&55 (20/35)&500&1.5-1.05-1.6&0.7-1.5&21 -  45\\
Night 3&2018-01-14&47 (20/27)&500&1.2-1.05-1.6&0.7-1.8&35 - 50\\
\hline
\end{tabular}
\tablefoot{\tablefoottext{a}{In parenthesis: spectra in- and out-of-transit, respectively.}
\tablefoottext{b}{Airmass at the beginning, centre, and end of transit.}\tablefoottext{c}{Order 56 contains the sodium feature.}}
\end{table*}

 \begin{figure}[htb]
\resizebox{\columnwidth}{!}{\includegraphics[trim=3.0cm 8.5cm 3.0cm 8.5cm]{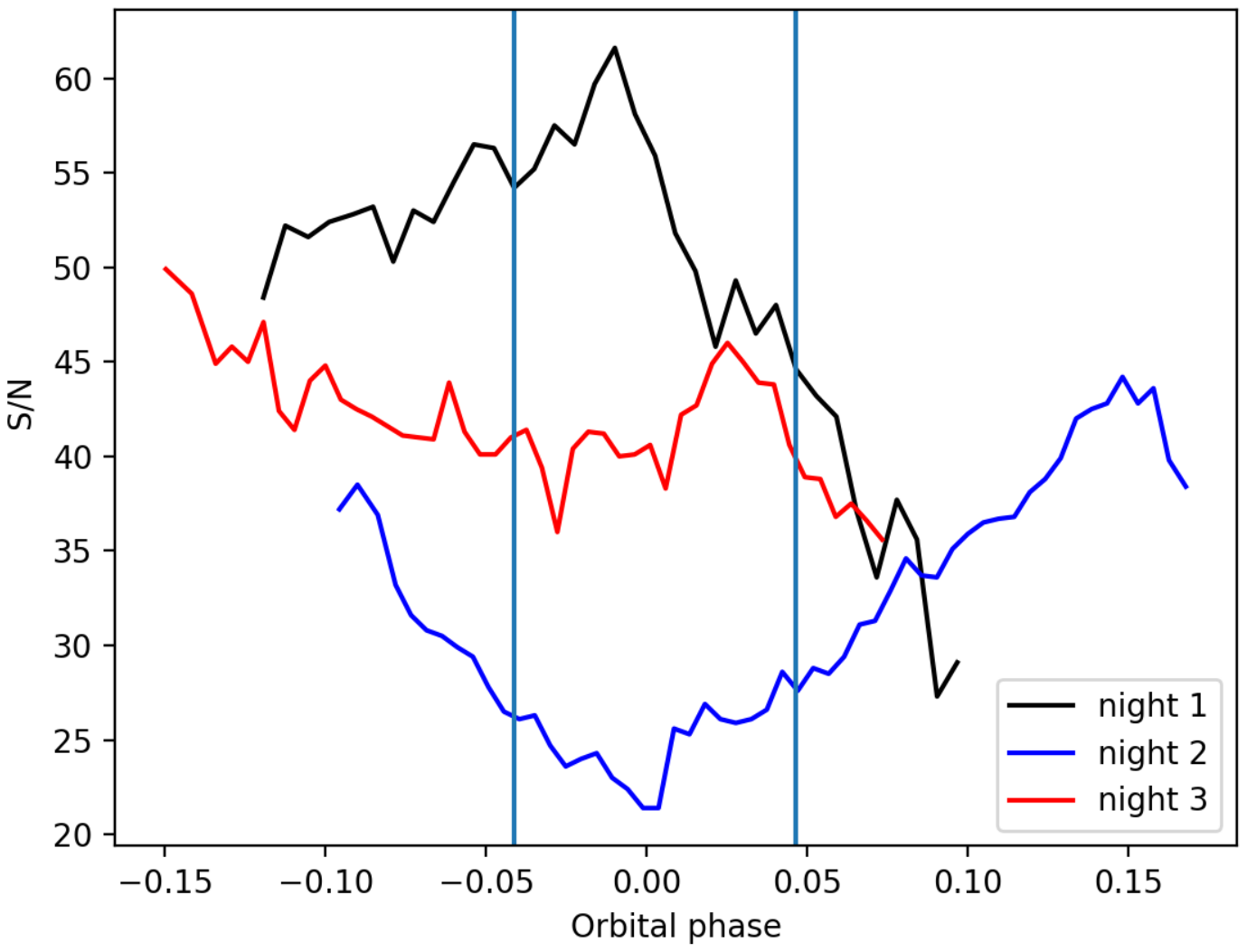}}
	\caption{SNR at the center of order 56 (near the location of the \ion{Na}{I} doublet) in all three nights as a function of orbital phase. The vertical lines indicate the start and end of the transit. The second night, in blue, shows markedly lower SNR during transit. For this reason, it is rejected during the focused analysis of the \ion{Na}{} doublet.}
	\label{fig:SNR}
\end{figure}

\noindent We used the HARPS Data reduction pipeline (DRS, version 3.8), of which the primary science products are the individually extracted Echelle orders from the 2D Echellogram, as well as blaze-corrected, stitched and resampled one-dimensional spectra. The wavelength solutions provided by the DRS are in air, in the rest frame of the observatory. The signal-to-noise ratio as recorded by the DRS at the center of order \#56 (which covers the sodium doublet) for each of the three nights are plotted in Fig. \ref{fig:SNR}.

\noindent Ground-based transmission spectroscopy at optical wavelengths of exoplanet atmospheres requires the correction of telluric contamination, caused by H$_2$O and O$_2$ absorption bands. Telluric contamination was corrected using the {\tt molecfit} package \citep{Smette2015,Ka15}, following previous work by e.g. \citet{Al17,Hoeijmakers2019} and \citet{Se19} for an application to the Na-D doublet.

\noindent We applied {\tt molecfit} to the one-dimensional spectra created by the pipeline to create a model of the telluric transmission spectrum over the entire wavelength range of HARPS for each spectrum in the time-series. These telluric models were then interpolated onto the wavelength solutions of each of the respective spectral orders and divided out, which yielded individually corrected Echelle orders. This correction is valid despite the fact that the Echelle orders are not blaze-corrected, because the telluric transmission spectrum and the blaze correction are both multiplicative operations. We visually inspected the spectra and found the correction to be effective down to the noise level at most wavelengths\footnote{The telluric correction is effective in all but the deepest O$_2$ lines, which are subsequently masked out in the analysis (see Section \ref{sec:masking}).}.

\noindent After telluric correction, the spectra were Doppler-shifted to place the host star in a constant rest-frame. To this end we performed two velocity corrections simultaneously: The Earth's velocity around the barycenter of the solar system, and the radial velocity of the star induced by the gravitational effect of the orbiting planet, leaving the stellar spectra at a constant velocity shift set by the systemic velocity of $\sim38$ \kms. This combined velocity shift constitutes the only re-interpolation of the extracted spectra during this cross-correlation analysis. Missing edge values were masked (see section \ref{sec:masking}). We proceeded to perform two independent analyses of the transmission spectrum of WASP-121: A focused analysis of the narrow waveband around the sodium D-lines (Sections \ref{sec:transspec} and \ref{sec:transspec_results}), and a cross-correlation analysis targeting metals and molecules with lines spread out over the waveband of HARPS (Sections \ref{sec:xcor} and \ref{sec:detections}).

\subsection{Narrow-band transmission spectroscopy}
\label{sec:transspec}
The in-transit spectra were divided by the mean of the out-of-transit spectra which constitutes a master out-of-transit spectrum. This implicitly removed the blaze, the stellar continuum flux and stellar absorption lines, and yields a time-series of transmission spectra in the rest-frame of the star. These were shifted to the planetary rest-frame using the known orbital parameters \citep{Bourrier2020} and averaged in time. The transmission spectra for the individual nights were then combined to create a transmission spectrum averaged over three transits. A more detailed description of this approach can be found in \cite{Wy15}. The normalisation processes, where all spectra are brought to the same flux level for a correct calculation of the in- and out-of-transit master, followed that of \cite{Se19}.

\subsection{Cross-correlation analysis}\label{sec:xcor}
Due to the high equilibrium temperature of this planet the transmission spectrum features absorption lines of atomic metals other than sodium, as already noted by \citet{Bourrier2019,Sing2019,Gibson2020,Cabot2020}. We use the cross-correlation method \citep{Snellen2010} to search for an ensemble of neutral atoms and ions, following the strategy applied in previous work \citep{Hoeijmakers2019}. \\

\noindent To implicitly retain information about the absolute flux recorded by the spectrograph, we have modified the analysis used by \citet{Hoeijmakers2018,Hoeijmakers2019} to be applied directly to the extracted spectral orders, as opposed to the blaze-corrected, stitched and resampled one-dimensional spectra that are also created by the data-reduction pipeline. Under the assumption that the noise is dominated by photon noise, retaining the absolute flux implicitly weighs spectral pixels according to their variance when performing the cross-correlation (see Section \ref{sec:cross_correlation}). \footnote{The same strategy is applied automatically by the DRS when it cross-correlates the data with standardized templates for the purpose of measuring the radial velocity of exoplanet host stars.}

\noindent First, the spectral orders (each with a width of $4096$ pixels) of the time-series were sorted in time and grouped into two-dimensional matrices of $4096 \times N$ values, where $N$ is the number of exposures obtained per night, equal to 63, 45 and 50 respectively. Because the wavelength-solution computed by the HARPS pipeline is constant during the time-series, this re-ordering did not require resampling of the extracted spectra.

\subsubsection{Masking and colour correction}\label{sec:masking}
We performed a two-step process to mask outlying flux values and spectral regions that are otherwise affected by artifacts. We first performed order-by-order sigma-clipping by computing a running standard deviation over a sub-region measuring $40 \times N$ values and flagging $5\sigma$ outliers from the mean as missing data. Secondly, we visually inspected all orders and select spectral channels (i.e. $1 \times N$ columns) where detector artifacts, regions with excessive noise (e.g. at the edges of the bluest orders), stellar residuals (e.g. in the cores of deep absorption lines) or imperfectly removed tellurics are apparent. These columns were also flagged. Both masking steps affected a total of 1.1\%, 2.7\% and 1.6\% of all the values in each of the three nights respectively.\\

\noindent Unless an entire spectral column is missing from the time-series, missing values would affect the evaluation of the cross-correlation function, effectively introducing a time-dependence of non-astrophysical origin\footnote{Points that are masked / set to NaN affect the evaluation of the sum over the cross-correlation template, $T$ in Eq. \ref{eq:ccv}. If the spectral channel is missing in only some of the exposures during the time-series, $T(i)$ becomes time-dependent. This would introduce an error when correcting one cross-correlation function by another, as is done when removing the stellar contribution. Columns with significant numbers of missing values were therefore masked out completely, while isolated missing values were interpolated over.}. We therefore distinguish between missing columns and isolated missing values. Missing columns were ignored, as these spectral channels do not contribute to the cross-correlation function at any time in the numerator of the cross-correlation function. Conversely, isolated missing values were interpolated over in the spectral direction. Columns of which over 20\% of values are missing were ignored from the analysis entirely. After masking, the spectral orders were duplicated and a model spectrum of the planet was injected into these data to enable model comparison (see Section \ref{sec:templates}). After model injection, these "contaminated" spectra were treated in the same way as the spectra in which no model was injected, following earlier practice \citep{Hoeijmakers2015,Hoeijmakers2018,Hoeijmakers2019}.\\

\noindent Because the broad-band continuum of the HARPS spectrograph varies slightly over the course of the observations, we performed a colour-correction of the spectral orders by normalizing the mean flux of each spectral order over wavelength to the mean flux of each spectrum over time. This enforces fixed ratios between the mean flux levels between the orders, consistent with the combined effect of the typical response function of the spectrograph, the transmission of the Earth's atmosphere and the intrinsic spectrum of the star. Because this normalisation step removes information about the average flux of the exposures as a function of time, we weighted the cross-correlation functions of the time series accordingly at the end of the analysis (see Section \ref{sec:cross_correlation}).

\subsubsection{Model spectra and cross-correlation templates}\label{sec:templates}
To create model spectra for purpose of model injection and as cross-correlation templates, we followed the same procedure as \citet{Hoeijmakers2019}: the atmosphere of the planet was assumed to be isothermal and in chemical and hydrostatic equilibrium, with elemental abundances corresponding to a fixed metallicity value. The chemical abundance profiles were computed using \texttt{FastChem} \citep{Stock2018}, and the radiative transfer was performed as described in \citep{Gaidos2017}, using opacity functions derived from line-lists provided by VALD and Exomol for atoms and molecules respectively \citep{Ryab2015,Tennyson2016}. We included opacity from all atomic neutrals and single ions with atomic numbers up to 40, due to diminishing abundance for higher atomic numbers. Although \texttt{FastChem} produces abundance profiles for hundreds of molecules, we only considered molecular opacity caused by TiO \citep{McKemmish2019}, VO \citep{McKemmish2016}, SH \citep{Gorman2019}, H$_2$O and H$_2$. This choice is partly motivated by diminishing abundances of larger molecules, but also by a sparsity in reliable line-list data for most molecules.\\

\begin{figure}[htb]
{\includegraphics[width=\linewidth]{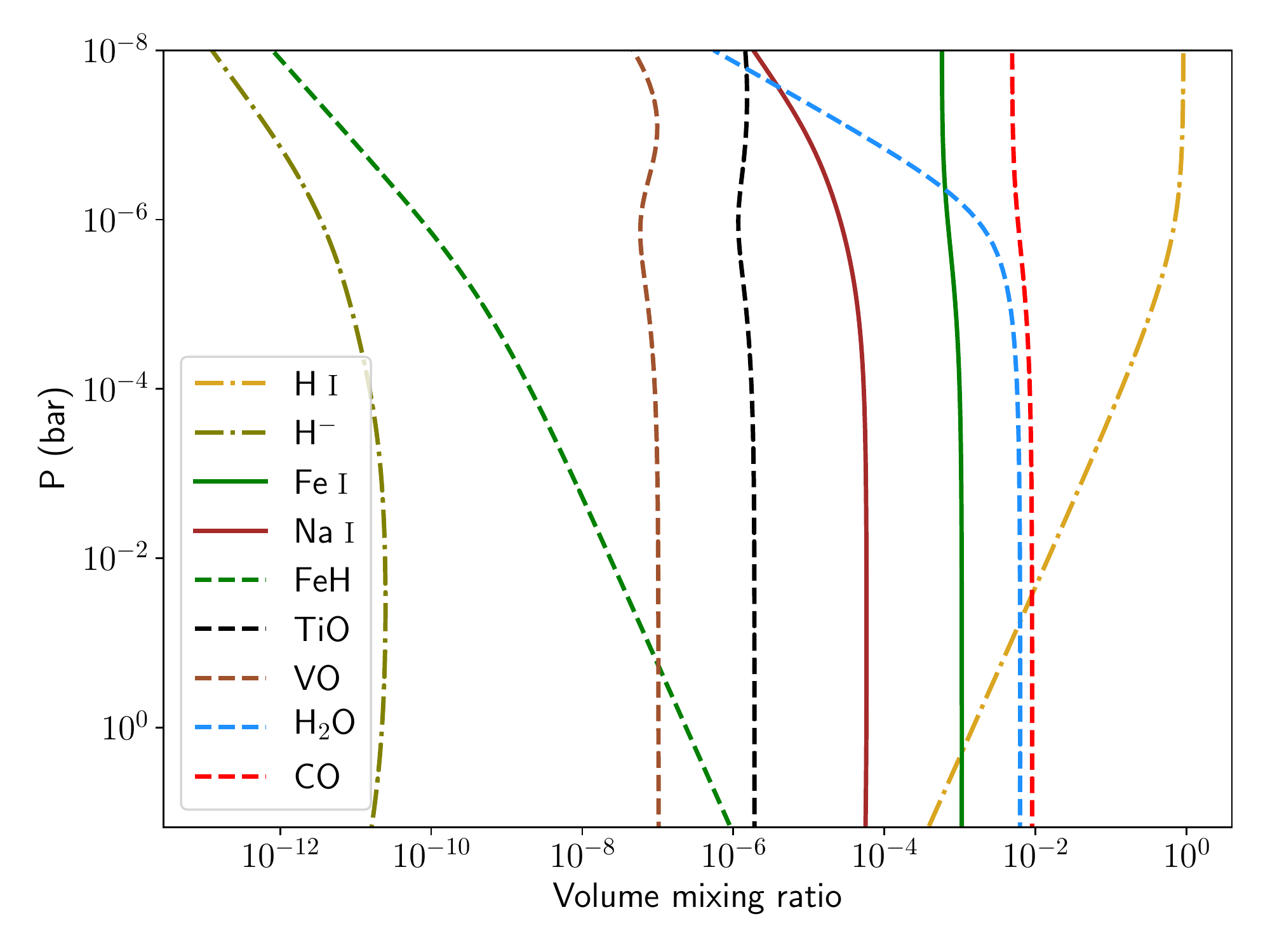}}
	\caption{Abundance profiles of selected species at a temperature of 2,000 K and 20$\times$ solar metallicity, as computed by \texttt{FastChem}. Solid lines correspond to atomic species, dashed lines to molecules, and dashed-dotted lines to atomic hydrogen and $H^-$. Transmission spectroscopy is principally sensitive to pressures below the milli-bar level \citep[e.g.][]{Kitzmann2018}.}\label{fig:chemistry_all}
\end{figure}

\begin{figure*}[htb]
\resizebox{\textwidth}{!}{\includegraphics[trim=0.0cm 0.0cm 0.0cm 0.0cm]{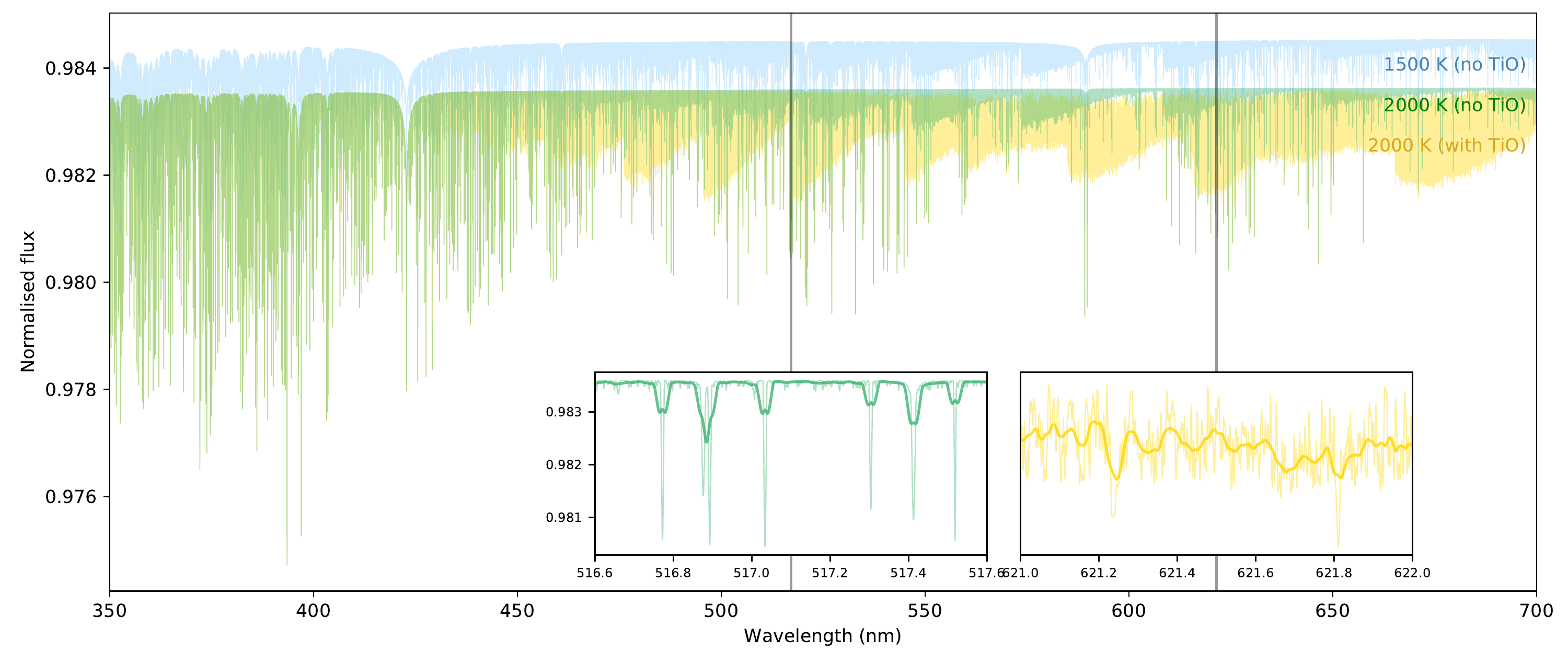}}
	\caption{Three of the four models of the transmission spectrum of WASP-121 b assuming chemical equilibrium, a metallicity of 20$\times$ solar, a temperature of 1,500 (blue) or 2,000 K, with (yellow) and without (green) TiO opacity, used for model injection and comparison. The inset panels show the models with (right) and without (left) the contribution of TiO opacity near 517 nm and 622 nm, both at the high native resolution of these models (light colour), as well as broadened to include the effects of the instrumental resolution and rigid body rotation assumed for the planet (dark colour). VO absorption bands are only evident when TiO opacity is removed (the bands that remain visible in the green and blue models near e.g. 550 nm and 575 nm), which would otherwise be masked by much stronger TiO bands. The fourth model (at 3,000 K) is not plotted here, but is shown separately in Fig. \ref{fig:STIS}.}
	\label{fig:models}
\end{figure*}

\noindent In this way, we produced a single spectrum for each absorbing species (including continuum absorption) assuming a temperature of 2,000 K, to be used as cross-correlation templates. For the purpose of model comparison, we additionally produced four model spectra assuming a temperature of 1,500, 2,000 or 3,000 K, a metallicity of 20$\times$ solar \citep[corresponding to the median value found by ][]{Evans2018}, and including the opacity of all species, but artificially disabling the effect of TiO opacity to account for the possible effects of TiO condensation, as listed in Table \ref{table:modelsoverview} and shown in Fig. \ref{fig:models}. The abundance profiles for the model at 2,000 K are shown in Fig.  \ref{fig:chemistry_all}.

 \begin{table}
\caption{Parameters of the models used for injection and comparison, three of which are shown in Fig. \ref{fig:models}. These models assume chemical and hydrostatic equilibrium, with elemental abundance ratios fixed to 20$\times$ the solar value \citep{Asplund2009}}
\label{table:modelsoverview}
\centering
\begin{tabular}{c c c c c c }
\hline
\hline
 \# & $T$ (K) & [Fe/H] & VO & TiO & SH    \\
\hline

 \hline
 1 & 1,500 & 20 & $\times$ &          & $\times$\\
 2 & 2,000 & 20 & $\times$ &          & $\times$\\
 3 & 2,000 & 20 & $\times$ & $\times$ & $\times$\\
 4 & 3,000 & 20 & $\times$ &          & $\times$\\
\hline
\end{tabular}
\end{table}

\noindent The four spectra that are used for model injection were convolved with a broadening kernel to simulate the effects of tidally locked rotation of the planet ($v_{\textrm{eq}} = 7.0$ \kms), the resolution of the spectrograph and the change of the radial velocity of the planet between the start and end of each exposure, following the approach by \citet{Brogi2016}. They were shifted to the average radial velocity of the planet during the exposure (assuming $v_{\textrm{orb}} = 221.1$ \kms and $v_{\textrm{sys}} = 38.043$ \kms), and multiplied with a normalized model of the transit light-curve\footnote{Using Ian Crossfield's Astro-Python Code: \url{http://www.mit.edu/\~iancross/python/}} before injection into each spectrum via multiplication.

\begin{figure*}
  \centering
    \includegraphics[width=0.88\linewidth]{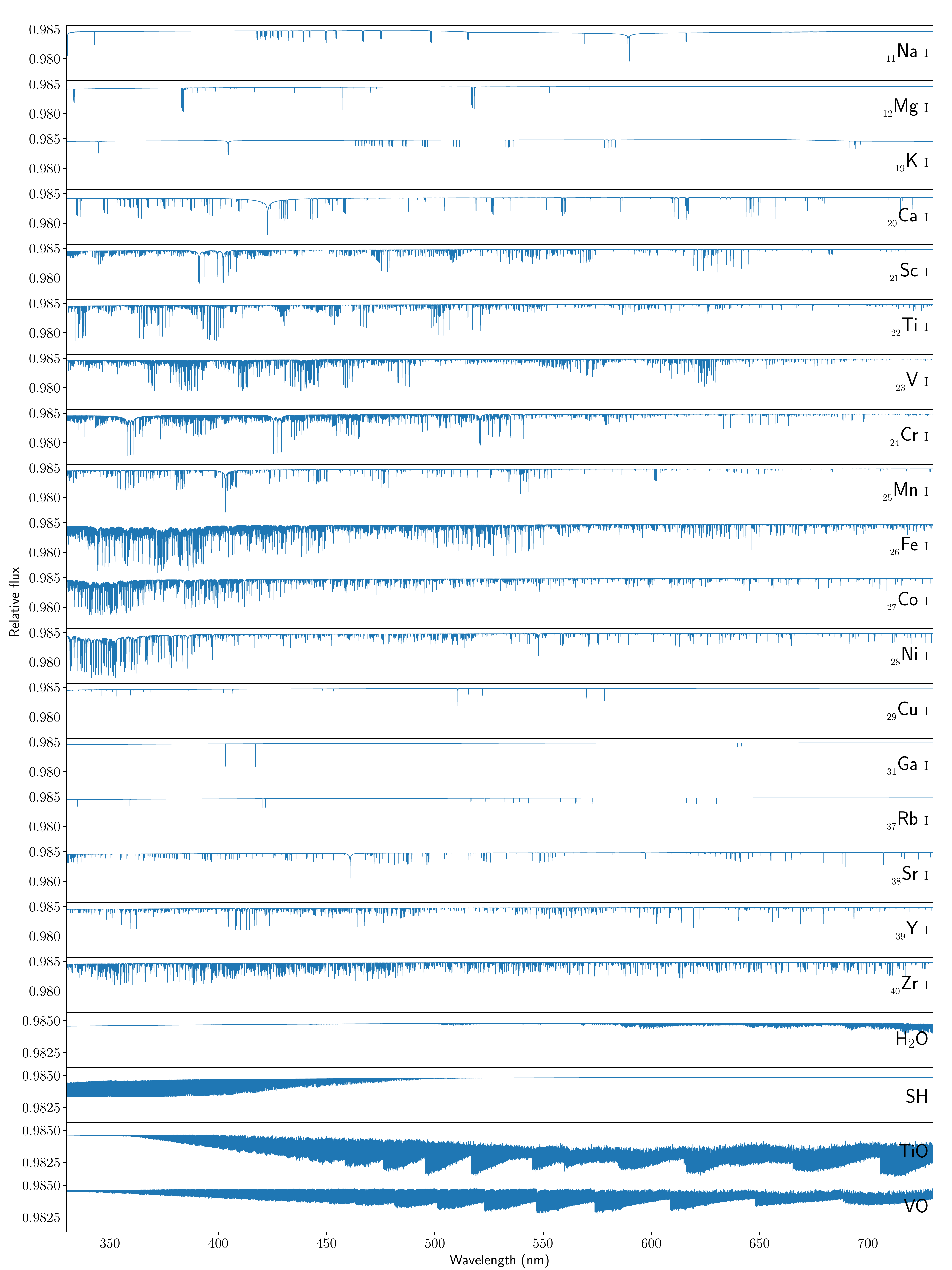}
	\caption{The cross-correlation templates introduced in this work prior to continuum subtraction and broadening, derived from model spectra of the atmosphere of WASP-121 b. The chemical abundances of the atoms assume chemical equilibrium with elemental abundances equivalent to a metallicity of $20\times$ solar and a temperature of $2,000$ K. The templates include all atoms with atomic number up to 40 for which line absorption is significant in the HARPS wavelength range (387.4 to 690.9 nm) under these conditions, as well as the H$_2$O, SH, TiO and VO molecules. SH is included as a byproduct of sulphur photo-chemistry, with a disequilibrium abundance profile artifically set to match the profile presented in \citet{Evans2018}. The profile is parametrised as having a VMR of $10^{-4}$
	for pressures higher than $10^{-5}$ bar, and zero at lower pressures. Note that for clarity, the scale of the y-axis of the molecular absorption spectra is a factor of 2 smaller than the atoms. Also note that the absorption lines of SH are expected to be significantly weaker than those of the atomic metals. Templates of ionic species (\ion{Ti}{II}, \ion{Cr}{II} and \ion{Fe}{II}) are adopted from the analysis of \citet{Hoeijmakers2019}, calculated for the atmosphere of KELT-9 b with a temperature of 4,000 K and solar metallicity.} \label{fig:templates}
\end{figure*}

\noindent The model spectra that are used as cross-correlation templates were continuum-subtracted by fitting a low-order polynomial. Any residual values that are smaller than $1\times 10^{-4}$ times the amplitude of the deepest spectral line in the waveband were clipped to zero, ensuring a constant continuum. Given that the deepest spectral lines in the optical have typical depths of $\sim1\times 10^{-3}$ times the flux of the star, this procedure affects only spectral lines at the $\sim1\times 10^{-7}$ level. Then, the template was convolved with a Gaussian kernel with a FWHM width of 2.72 \kms, matching the resolving power of the spectrograph. This ensured that the template is never undersampled when interpolated onto the wavelength grid of the observed spectra \footnote{Broadening the template by an additional 2.72 \kms causes artificial broadening of the resulting cross-correlation function by the same amount, which needs to be taken into account when interpreting the width of the cross-correlation function. However, because the linewidth of the planet is typically significantly greater (due to diurnal rotation of the planet, amounting to $v_{\textrm{eq}} = 7.0 $ \kms in the case of WASP-121 b), and because broadening terms add in squares, the contribution of this broadening term on the end result is small.}. All templates developed for this analysis are shown in Fig. \ref{fig:templates}. \\

\noindent The above templates do not contain lines by ionized metals because thermal ionization is insignificant at 2,000 K. To test for the presence of absorption lines by ions, we used templates computed for the transmission spectrum of KELT-9 b \citep{Hoeijmakers2019} at a temperature of 4,000 K and solar metallicity. These templates are publicly available via the CDS \footnote{Via anonymous ftp to \url{cdsarc.u-strasbg.fr} (130.79.128.5) or via \url{http:// cdsarc.u-strasbg.fr/viz-bin/qcat?J/A+A/627/A165}}.\\

\subsubsection{Cross-correlation}\label{sec:cross_correlation}
 The cross-correlation operation (Eq. \ref{eq:ccv}) was applied to the extracted orders without correcting the blaze function nor the stellar absorption lines. Therefore, the correlation coefficients $c(v,t)$ constitute flux measurements as the number of photon counts registered by the detector (averaged over the time series due to the colour-correction described above). The functional form of the cross-correlation coefficient $c(v,t)$ as an averaging operator is:

\begin{equation}\label{eq:ccv}
c(v,t) = \sum_{i=0}^{N_x}x_i(t) T_i(v),
\end{equation}

 where $x_i(t)$ are each of the spectral points in all the echelle orders of the spectrum obtained at time $t$, $T_i(v)$ are the corresponding values of the template Doppler shifted to a radial velocity, $v$. $T$ takes on non-zero values inside spectral lines of interest, and is normalized such that $\sum_{i=0}^{N}T_i(v) = 1$. The sum includes the spectral points of all echelle orders, eliminating the need for explicit order-by-order weights as applied in earlier works \citep{Hoeijmakers2018,Hoeijmakers2019}, as well as computationally expensive filtering steps applied to the full high-resolution spectrum. The transit-depth associated with the atmosphere of planet is obtained from $c(v,t)$ by dividing all the cross-correlation functions of the time-series by their out-of-transit average, reminiscent of studies that use the cross-correlation functions produced by the pipeline directly \citep{Bourrier2019,Ehrenreich2020}.\\

\begin{figure*}
  \centering
    \includegraphics[width=0.9\textwidth]{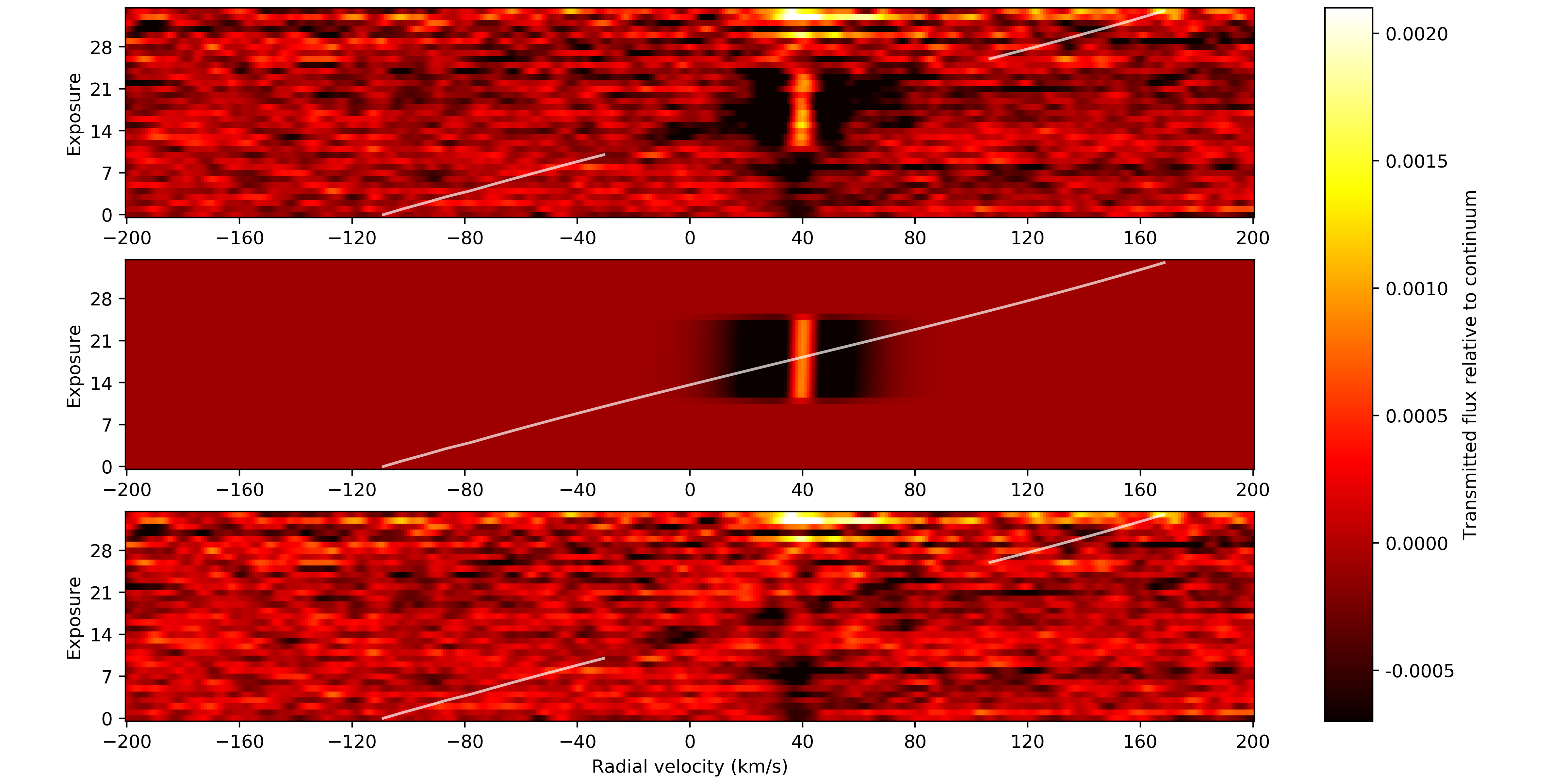}
	\caption{The subtraction of the Doppler shadow and the emergence of the signature of the exoplanet atmosphere, following the radial-velocity curve indicated by the white line. \textbf{Top panel}: Raw cross-correlation matrix of the first night of observations with the \ion{Fe}{I} template. During the transit, the Doppler shadow emerges as the near-vertical structure around $40$ \kms, because the planet is on a strongly misaligned orbit \citep{Delrez2016,Bourrier2020}. \textbf{Middle panel:} Best-fit model of the Doppler Shadow described as a two-component Gaussian. \textbf{Bottom panel:} Residuals after subtracting the best-fit model from the raw cross-correlation matrix, revealing the signature of the planet atmosphere. Residuals remain in the stellar line core before and after the transit event, but these do not affect the signature of the planet atmosphere, as the latter is constructed by considering in-transit spectra only. The signature of the planet atmosphere appears as the dark slanted feature indicated by the white line. Note that although the sign of the absorption is negative in this figure, the sign flipped further in the analysis to denote absorption, notably in Fig. \ref{fig:detections} and \ref{fig:molecules}.}
	\label{fig:shadow}
\end{figure*}

\noindent Like in previous work, we subtracted an empirical model of the Doppler shadow that is caused by the obscuration of part of the rotating stellar disk during the transit of the planet. We used the double- Gaussian model introduced by \citet{Bourrier2018} for the M dwarf GJ 436, which consists of the sum of two Gaussians with a positive central core and negative side-lobes (see Fig. \ref{fig:shadow}). The resulting model multiplied by a scaling factor is subtracted from the cross-correlation function obtained for each template, minimizing the sum of the squared residuals while ignoring cross-correlation values at velocities that coincide with the expected radial velocity of the planet. We also applied a high-pass filter with a width of 70 \kms that removes residual broad-band variations that were not already removed during the colour-correction stage \citep{Hoeijmakers2018,Hoeijmakers2019}.

\noindent Finally, the cross-correlation functions were shifted to the expected rest-frame velocity of the planet (based on its assumed orbital velocity and ephemeris), weighted according to the mean flux in their corresponding exposures (as already mentioned in Section \ref{sec:masking}), masked according to whether they are expected to be in-transit (1.0) or out-of-transit (0.0), and co-added. This yields a time-averaged, one-dimensional cross-correlation function in the rest-frame of the planet, for each of the three nights. These were finally averaged again without application of weights. This entire analysis is illustrated in Fig. \ref{fig:analysis_steps}.

\section{Results and discussion}\label{sec:results}

\subsection{Transmission spectra}
\label{sec:transspec_results}

\begin{figure*}[htb]
 \centering
    \includegraphics[width=0.9\textwidth]{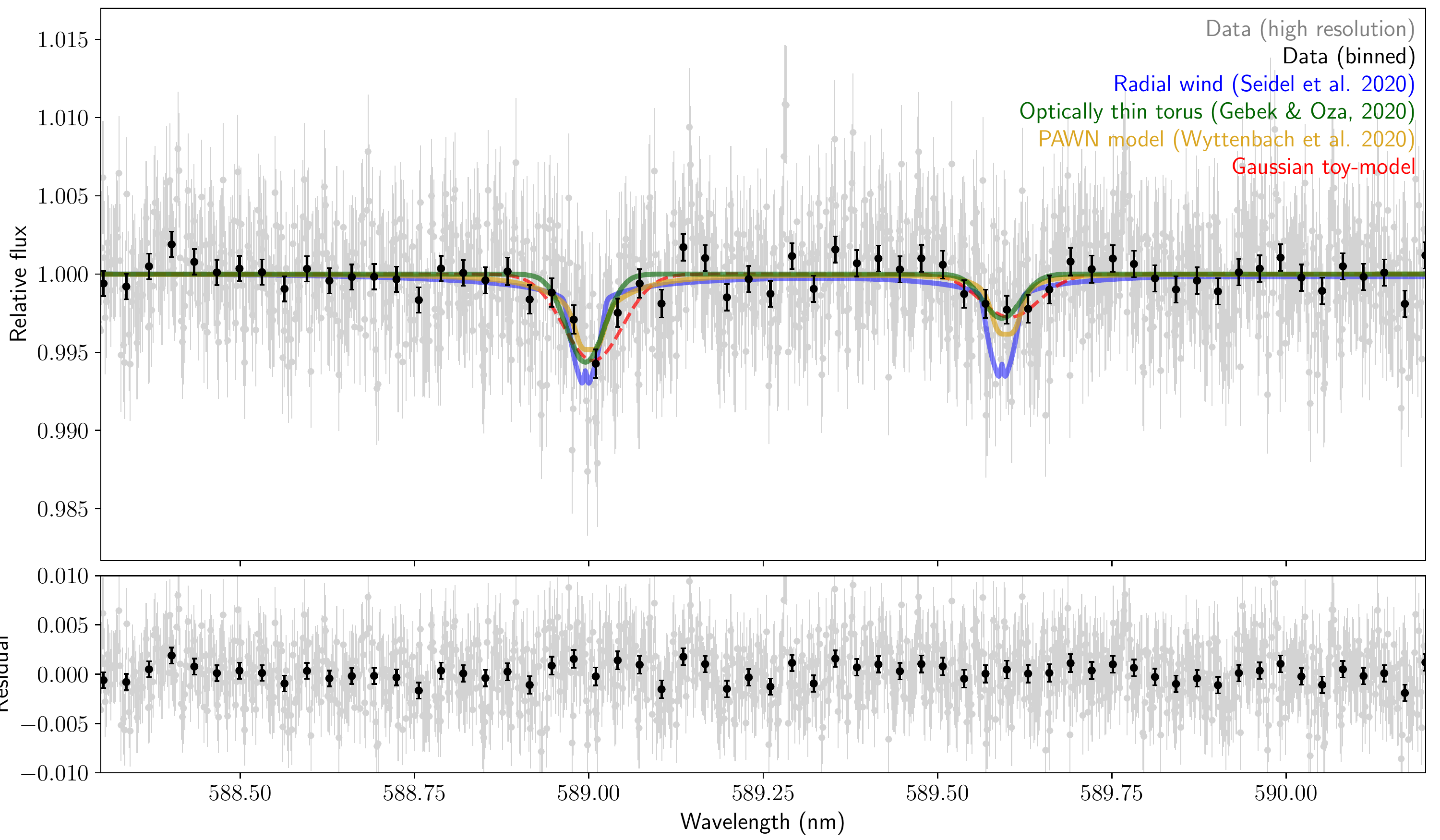}
	\caption{Transmission spectrum of WASP-121 b at the location of the \ion{Na}{I} D-doublet. The spectrum was constructed from the observations obtained during night one and night three, as night two was ruled out as a spurious signal. The top panel shows the transmission spectra at the native resolution of HARPS in grey, binned by 20 for better visibility in black. The transmission spectrum corresponding to the best-fit analytical model (see Sections \ref{sec:sodiumresults} and \ref{app:sodium}) is shown in red, along with the residual of the data after subtraction in the bottom panel. Blue, green and yellow lines depict two forward-models based on a radially outflowing wind \citep{Seidel2020}, an optically thin torus \citep{Oza2019,Gebek2020} and a hydrodynamically escaping envelope \citep{Wyttenbach2020}. The latter two reproduce the significant difference in depth between the D1 and D2 lines.}
	\label{fig:NaSpectrum}
\end{figure*}

Fig. \ref{fig:NaSpectrum} shows the transmission spectrum of WASP-121 b in Echelle order ($\#56$) of the sodium doublet in the planetary rest frame. The relative depths and detection levels are calculated by fitting a Gaussian to both peaks taking into account the propagated error of the individual spectral samples. The relative depths of the \ion{Na}{I} D-lines are hence obtained directly from the differential transmission spectrum without resorting to pass-bands in the blue and red arms of the sodium doublet as was done by \citet{Wy15,Wyttenbach2017,Se19}. These are summarised in Table \ref{table:absorbnightly}.

\subsubsection{Assessment of data quality}
\label{sec:dataquality}
Each transit in this data-set is possibly contaminated by spurious occurrences such as stellar spots, instrumental effects or adverse observational conditions, leading to false-positive detections that do not stem from the exoplanet transit itself. To rule out a false-positive detection due to these systematic errors and estimate their likelihood, we performed a bootstrap (or empirical Monte-Carlo, EMC) analysis. We follow the approach in \cite{Re08}, where sub-samples of the dataset are randomly selected to be fed to the analysis, creating randomised instances of the transmission spectra. \cite{Re08} explores three scenarios: taking sub-samples only from the in-transit spectra (in-in), only from the out-of-transit spectra (out-out), or from both the in-transit spectra and out-of-transit spectra respectively (in-out). If the detected \ion{Na}{I} signature is caused by absorption in the planet atmosphere, we expect to find no signal when dividing the the in-in and out-out spectra with each other. More detailed examples of this technique can be found in \cite{Wy15,Se19}.\\

\noindent Apart from assessing the reliability of the detection, the computed distributions serve the secondary purpose of estimating an upper boundary on the false-alarm probability of the signal, accounting for varying observation conditions and instrumental effects \citep{Re08}. The standard deviation of the out-out distribution, which is unaffected by the planetary atmosphere, is used as the error on the measured absorption depth in \cite{Re08}. We weigh the standard deviations by the square root of the ratio of out-of-transit spectra to the overall number of spectra taken during each night, thus accounting for the biased sample selection in the out-out scenario \citep{Wy15}. This has been explored by \cite{As13}, who suggest the standard deviation of the in-out scenario as an alternative. To eliminate any influence from the planetary signal, we follow \cite{Wy15} in this analysis.

 \begin{figure*}[htb]
 \centering
\resizebox{0.9\textwidth}{!}{\includegraphics[trim=0.0cm 6.5cm 0.0cm 6.0cm]{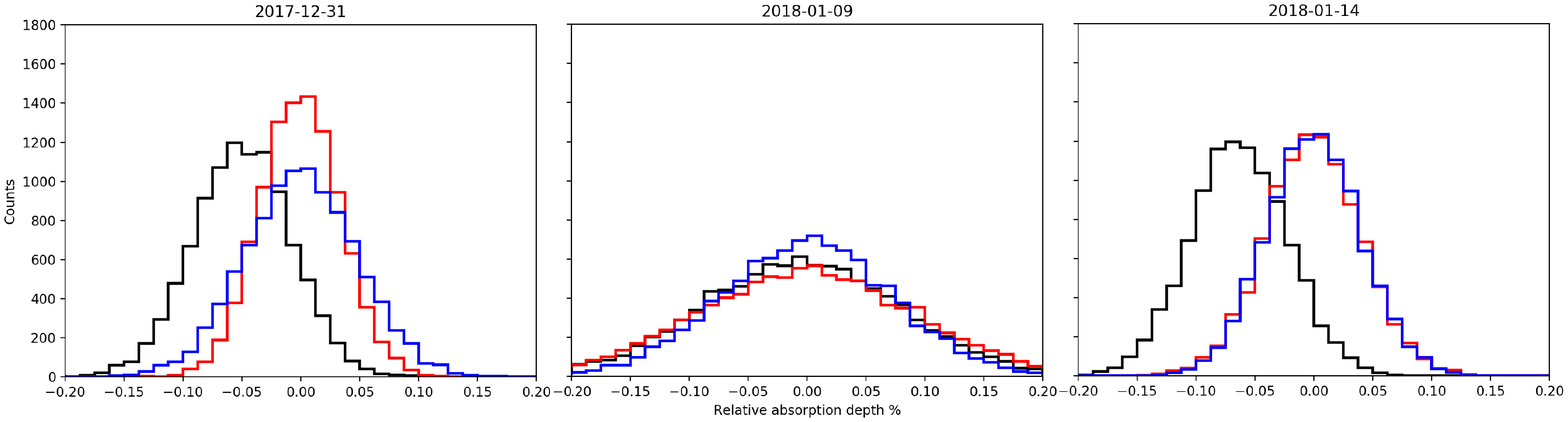}}
	\caption{Distributions of the bootstrap analysis for the $12$ \AA \xspace central passband. As expected the `in-in' (red) and `out-out' (blue) distributions are centred around zero (no planetary detection) and the randomised `in-out' distribution shows a detection (black) for both night one and night three. Night two shows wide variability in all three scenarios, with the 'in-out' distribution also centered at 0.}
	\label{fig:MCMCNa}
\end{figure*}

\noindent The results of the bootstrap analysis for each of the three nights can be found in Fig. \ref{fig:MCMCNa} for the spectral order of the sodium doublet as an example. We performed the bootstrap $10,000$ times for each scenario in each night, often enough for the standard deviation to not change significantly when the iterations are increased. The in-in scenario is shown in red, the out-out scenario in blue and the in-out scenario in black. As predicted, in all three nights the in-in and out-out scenarios are centered at $0$.
For both night one and night three, the sodium detection can clearly be seen at the same detection level as in the transmission spectra, with the black histogram offset with respect to 0.  However, night two shows no detection of sodium in transmission in the EMC. The signal to noise ratio (SNR) in night two for the order containing the sodium doublet (see Fig. \ref{fig:SNR}) shows a significantly lower SNR for the in-transit spectra (SNR $\sim 20$) compared to the out-of-transit spectra (SNR $\sim 45$), see Table \ref{table:nightoverview} and Fig. \ref{fig:SNR}. This indicates that the feature in the transmission spectrum of night two is likely a false-positive due to systematic noise and not dominated by Gaussian noise. We therefore decided to exclude night two from the transmission spectrum analysis. \\

\noindent Additionally we found residuals of stellar activity in the obtained spectra for all three nights, despite our correction for the stellar spectral lines (see Section \ref{sec:transspec}), similar to what has been observed by \citet{Bourrier2020}. To correct this contamination, we masked a window of 1.5 \kms around the center of the stellar sodium line in the stellar rest-frame, containing the entire stellar line core, before shifting into the planetary rest frame.

\subsubsection{Asymmetry in the sodium doublet?}\label{sec:sodiumresults}
\noindent Both sodium lines appear to be significantly broadened compared to what is expected from hydrostatic models, with best-fit Gaussian FWHM values of $66 \pm 17$ and $47.9 \pm 6.6$ \kms. Significant broadening of either or both of the \ion{Na}{I} lines has been observed in other hot Jupiters, with FWHM widths in excess of 20 \kms observed for HD 189458 b, WASP-49 b, WASP-76 b and WASP-52 b \citep{Wy15,Wyttenbach2017,Se19,Chen2020}. \citet{Seidel2020} introduce a physical model (\texttt{MERC}) that aims to explain this broadening by invoking radially outflowing winds with speeds near the escape velocity of the planet, as could be expected in the presence of strong atmospheric escape. A forward-model with a vertical wind speed of 30 \kms, an isothermal, hydrostatic\footnote{The \texttt{MERC} model assumes a hydrostatic density profile while adding a spherically-symmetric velocity component (i.e. radially outward from the planet surface) to the absorption lines of the absorbing species.} profile of 3,000 K and a uniform \ion{Na}{I} abundance of $\log X_{\textrm{Na}} = -2.6$ is superimposed on the observed transmission spectrum in Fig. \ref{fig:NaSpectrum}. This model qualitatively reproduces the strength and width of the stronger D2 line, but we note a difference between the strengths of the D1 and D2 lines ($2.7\times10^{-3}$ and $5.6\times10^{-3}$ respectively, with a ratio of $2.10\pm0.62$), the former being significantly over-predicted by the \texttt{MERC} model. This difference has also been observed in a recent, independent analysis by \citet{Cabot2020}.\\

\noindent Given the limited signal-to-noise of the detected sodium doublet, and a possibility for systematic errors (one of the three nights of observation used in this analysis has been discarded), it is possible that the observed difference between the D-lines includes uncorrected systematic effects, and we propose that future observations of the sodium doublet in WASP-121 b would be valuable to confirm this difference. If the D2 line is indeed significantly deeper than the D1 line, we propose the following astrophysically motivated hypothesis.

\subsubsection{An optically thin envelope}
\noindent Assuming a temperature of 2,000 K a mean particle weight of 2.3 u and an atmosphere that is in hydrostratic equilibrium, the depths of the D1 and D2 lines can be expressed in terms of the atmospheric scale height $H$, corresponding to $(14.1\pm3.8) H$ and $(29.6\pm3.4) H$, respectively, with a difference of $(15.5\pm5.1) H$. Isothermal hydrostatic theory predicts that the difference between the depths of the two sodium lines in units of $H$ is proportional to logarithm of the oscillator strengths \citep{Brown2001,Lecavelier2008,Benneke2012,Heng2015,Heng2017}, i.e.:

\begin{equation}
    R(\lambda_{\textrm{D}2})-R(\lambda_{\textrm{D}1}) = H \ln \frac{\kappa_{\textrm{D}2}}{\kappa_{\textrm{D}1}} = H \ln \frac{f_{\textrm{D2}}}{f_{\textrm{D1}}} = 0.69 H
\end{equation}

\noindent Therefore, an altitude difference of $15$ scale heights would not be expected to be observed, and line ratios significantly greater than unity are indeed not typically observed in other hot Jupiters with the exceptions of WASP-69 b \citep{Casasayas-Barris2017} and possibly HD 189733 b \citep{Wy15}.\\

\noindent Under the assumption of an isothermal hydrostatic structure \citep{Brown2001}, the number density profile of \ion{Na}{I} particles is modelled as an exponential function of altitude. Depending on the opacity $\kappa$, the optical depth crosses a $\tau \sim 1$ surface at some altitude, which is approximately equal to the transit radius at wavelength $\lambda$ \citep{Lecavelier2008,Heng2017}.\\

\noindent Recently, \citet{Gebek2020} showed that the Na-D line ratio in the transmission spectrum approaches a value of 2 when the absorbing \ion{Na}{I} gas is optically thin ($\tau \ll 1$) across both absorption line cores: In the optically thin limit, the line depth ratio is proportional to the the ratio of oscillator strengths \citep{Draine2011}, i.e. $\sim 2$ for the \ion{Na}{I} D-doublet. In this scenario, the absorption lines can be used to measure the column density of absorbing sodium atoms \citep{Draine2011}.

\noindent In Appendix \ref{app:sodium} we present an analytical model of the transmission spectrum of a transiting, optically thin \ion{Na}{I} cloud associated with WASP-121 b assuming Gaussian line shapes. We demonstrate the efficacy of this scenario by deriving  the total mass of \ion{Na}{I} needed to produce the observed absorption lines in the transmission spectrum of WASP-121 b as a function of the radii of the planet ($R_p$) and the star ($R_*$) and the observed line depths ($W$) and widths ($\Delta v$):

\begin{equation}\label{eq:massdensity_main}
    M_{\rm Na} = \pi R_*^2 \left( 1 - \frac{R_p^2}{R_*^2} \right) \frac{m_e c}{\sqrt{\pi} e^2}\frac{\sqrt{2} \Delta v}{f \lambda_0} m_\textrm{Na} W(\lambda_0).
\end{equation}

\noindent Through equation \ref{eq:massdensity_main}, each line provides a best-fit value for $M_{\rm Na}$ given its Gaussian fit parameters (see Table \ref{table:absorbnightly}) which yield an error-weighted average of the total mass column of \ion{Na}{I} of $M_{\rm Na} = (3.6 \pm 0.6) \times10^{10}$ g and $\delta v = 21.37 \pm 2.61$ (FWHM = $50.32\pm6.15$). This result is independent of the size or shape of the absorbing cloud, as long as it is smaller than the extent of the stellar disk and as long as it is indeed optically thin for a cloud of this size (a smaller cloud needs a larger optical depth to produce a given line depth and vice versa). When assuming that the cloud takes the (projected) shape of a ring around the planet with an outer radius of $2R_p$ and an inner radius of $R_p$ (see Fig. \ref{fig:schematic}, $A_c = 1.48\times10^{21}$ cm$^2$) and filling in Eq. \ref{eq:linedepth}, we obtain optical depths in the D-line cores of $\tau_{\textrm{D1}} = (5.7\pm1.6)\times10^{-2}$ and $\tau_{\textrm{D2}} = 0.120\pm0.016$, which means that the optically thin approximation is justified for a cloud that is contained within the Roche radius of the planet \citep[$R_L\sim2R_p$,][]{Sing2019}.\\

\noindent The presence of \ion{Na}{I} at high altitudes may be consistent with the observation that WASP-121 b possess an extended outflowing envelope \citep{Cabot2020} that contains metals \citep{Sing2019}. As shown above, if the number density of \ion{Na}{I} is low but extended enough, such an outflow may be largely optically thin throughout the D-line cores. We fit the observed sodium lines in WASP-121b by using the \texttt{PAWN} model developed by \citet{Wyttenbach2020}. This 1-D model retrieves structural parameters of a dynamically escaping outflow from high-resolution transmission spectra. We tested an isothermal hydrodynamic atmospheric structure with a transonic Parker wind solution. The atmosphere is considered in chemical equilibrium and the number densities of the electronic energy levels of sodium are kept free to allow for departures from thermodynamic equilibrium (non-LTE). The best solution is found for a temperature of $T=3700^{+600}_{-300}$ K, and a mass loss rate of $\log_{10}\dot{M} = 9.0\pm2.8$ g s$^{-1}$, and is shown in Fig. \ref{fig:NaSpectrum}. This model exhibits significant asymmetry in the line-depths, indicating that the optical depth is small at most altitudes. If the \ion{Na}{I} doublet is indeed formed in a hydrodynamic outflow,  these observations provide further evidence that atmospheric escape could play a non-negligible role in shaping the evolution of WASP-121b.\\

\noindent Scaling the estimated photo-ionization rate published by \citet{Huebner2015}, \citet{Oza2019} calculate the typical lifetimes of neutral sodium at low pressures in the upper atmospheres of a sample of hot Jupiters and find that these are on the order of minutes, implying that an extended optically thin cloud of neutral sodium is not stable against photo-ionization. \citet{Gebek2020} present a model of \ion{Na}{I} envelopes around hot Jupiters, based on the empirical density profile of the \ion{Na}{I} torus that surrounds Jupiter. Sputtering and charge exchange processes feed this envelope with fast ($\sim 10 - 100$) \kms neutral sodium atoms, replenishing atoms that are lost due to photo-ionization while simultaneously providing significant non-thermal line broadening \citep[][and references therein]{Wilson2002}. Fig. \ref{fig:NaSpectrum} includes a forward-model of the resulting transmission spectrum assuming a total \ion{Na}{I} mass of $7.2\times10^{10}$ g and a mean \ion{Na}{I} velocity of $\left<v\right> = 22$ \kms. Like the optically thin toy-model, this model is also able to reproduce line depth ratios significantly greater than unity. \\

\noindent Regardless of the physical origin of the observed \ion{Na}{I} D-line ratio in the present observations, verification of significant \ion{Na}{I} line broadening and depth-ratios different from unity as observed in the transmission spectra of some hot Jupiters may provide important clues to the structure and origin of their envelopes, and open the doors to a better understanding of these environments via ab initio modeling \citep[e.g.][]{Steinrueck2019}, \citep{Showman2019} or data driven exploration \citep[e.g.][]{Waldmann2015,Fisher2019,Oza2019,Seidel2020,Gebek2020}.

\begin{table*}
\caption{Relative depth and detection levels of atmospheric sodium on WASP-121 b for each of the nights and all nights combined. The second night was rejected as a spurious signal and is not taken into account.}
\label{table:absorbnightly}
\centering
\begin{tabular}{c | l c c c }
\hline
\hline
Line&&Date& Line depth ($\times10^{-3}$) & $\sigma$  \\
\hline
&Night 1&2017-12-31& $3.50\pm0.97$&$3.61$\\
D$_1$&Night 2\tablefootmark{a}&2018-01-09&--&--\\
&Night 3&2018-01-14&$2.49\pm1.05$&$2.37$\\
\hline
&Combined\tablefootmark{a}&&$2.66\pm0.71$&$3.7$\\
\hline
\hline
&Night 1&2017-12-31& $6.62\pm0.88$&$7.52$\\
D$_2$&Night 2\tablefootmark{a}&2018-01-09&--&--\\
&Night 3&2018-01-14&$4.66\pm0.95$&$4.91$\\
\hline
&Combined\tablefootmark{a}&&$5.59\pm0.65$&$8.60$\\
\hline
\end{tabular}
\tablefoot{\tablefoottext{a}{Excluding night two, where we have shown that the apparent detection is likely spurious.}}
\end{table*}

\subsection{Cross-correlation analysis}\label{sec:detections}
\begin{figure*}
    \includegraphics[width=1.0\textwidth]{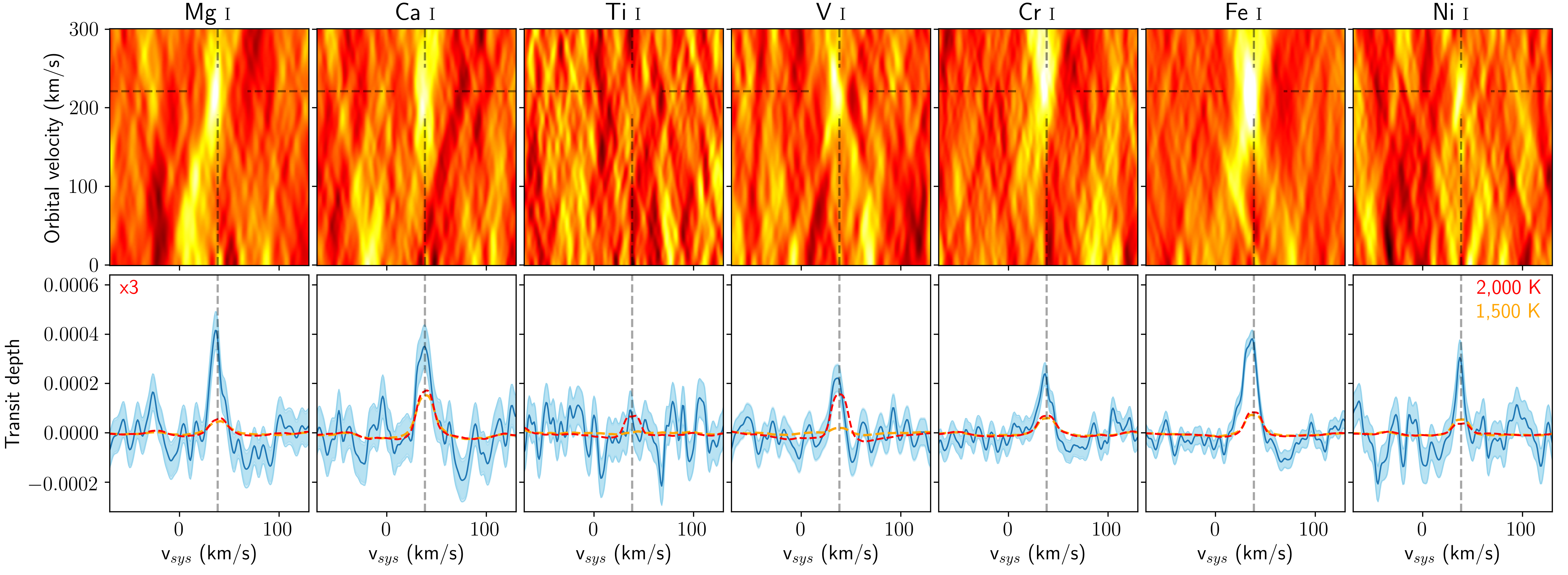}
	\caption{Summary of co-added cross-correlation functions in velocity-velocity space (top panels) and in the planet rest-frame (bottom panels), for various species. Dashed lines indicate the velocities at which the signature of the planet is expected to occur, given the known systemic velocity and orbital parameters of the planet. Red and yellow lines are cross-correlation functions obtained by injecting two different models into the data prior to cross-correlation. The cross-correlation functions of all surveyed species are provided in Fig. \ref{fig:all_ccfs}. Note that although the sign of the absorption is positive in this Figure, the sign was flipped earlier in the analysis to denote the flux received from the system, notably in Fig. \ref{fig:shadow} and \ref{fig:NaSpectrum}. Also note that the scaling of the vertical axis of the cross-correlation function of \ion{Mg}{I} is multiplied by a factor of three, because the line of \ion{Mg}{I} is significantly deeper than the other atoms.}	\label{fig:detections}

    \includegraphics[width=1.0\textwidth]{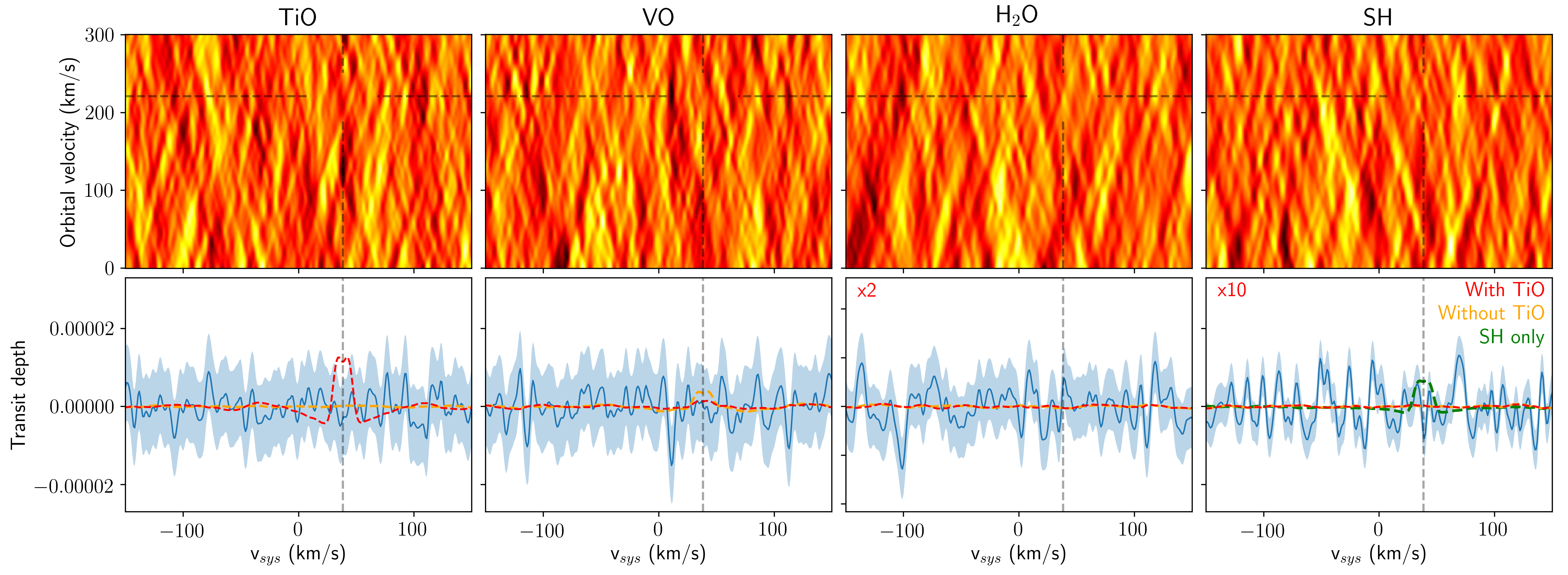}
	\caption{Similar to Fig. \ref{fig:detections}, but for molecular species. No molecules are detected using the cross-correlation technique. The y-axis values of the 1D cross-correlation functions of H$_2$O and SH are increased by factors of 2 and 10 respectively, to accommodate for their larger scatter. The injected model of SH (green) includes only continuum opacity and line opacity by SH at a temperature of 2,000 K.}\label{fig:molecules}

    \includegraphics[width=1.0\textwidth]{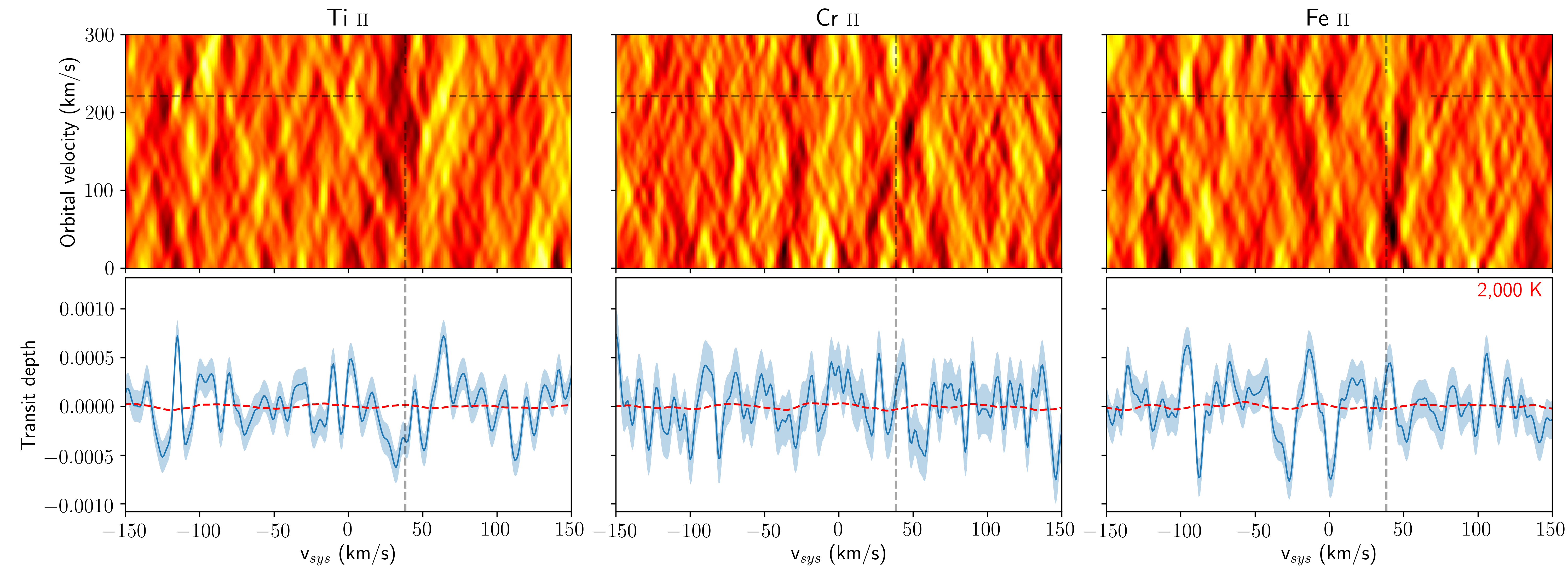}
	\caption{Similar to Fig. \ref{fig:detections}, but for ions. No ions are detected using the cross-correlation technique.}\label{fig:ions}
\end{figure*}

Fig. \ref{fig:detections} summarises the cross-correlation functions of the most important atomic species surveyed in this analysis, showing detections of \ion{Na}{I}, \ion{Mg}{I}, \ion{Ca}{I}, \ion{V}{I}, \ion{Cr}{I} and \ion{Fe}{I}, as well as a detection of \ion{Ni}{I} and a non-detection of \ion{Ti}{I}. Similarly, Figures \ref{fig:molecules} and \ref{fig:ions} show the cross-correlation functions of TiO, VO, H$_2$O, SH, \ion{Ti}{II}, \ion{Cr}{II} and \ion{Fe}{II}, none of which are detected. We fit the detected species with Gaussian profiles to determine the line positions, depths and widths. Because the cross-correlation is performed in steps of 1 \kms, and the templates were broadened to a FWHM of 2.7 \kms (see section \ref{sec:templates}) neighbouring points in the cross-correlation function are correlated with each other up to a range of $\sim3-4$ steps. When fitting Gaussian profiles, we therefore fit only to each fourth sample of the cross-correlation function, following the approach by \citet{CC2010b}. These fitting results are shown in Table \ref{table:detections}.\\

\noindent We performed two types of bootstrap analysis to confirm the strength and confidence level of each. The first bootstrap method is similar to the one applied to the lines of the sodium doublet in the transmission spectrum, comparing the distributions obtained by randomly selecting subsets of the in and out-of-transit spectra (Section \ref{sec:dataquality}).
The second bootstrap method uses the wide velocity range over which the cross-correlation is performed to construct a distribution of the systematic noise at Doppler velocities away from the main planetary and stellar signals (i.e. near $v_{\textrm{sys}}$. This is achieved by randomly Doppler-shifting each of the cross-correlation functions of the 2D time-series (bottom panel of Fig. \ref{fig:shadow}) of each species before time-averaging the series, and measuring the distribution of Gaussian fit parameters that occur at random. Both bootstrap analyses are detailed further in Appendix \ref{app:bootstrap}.\\

\begin{table*}
\caption{Gaussian fits to the detected species shown in Fig. \ref{fig:detections}. The line amplitude (A) is equivalent to the fractional area of the star that is obscured. In the last column, the line amplitude is compared to the amplitude obtained after injection of a model spectrum containing opacity of all species minus TiO, with a temperature of 2,000 K (model \#2 in Fig. \ref{fig:models} and Table \ref{table:modelsoverview}). Note that the line-amplitude of \ion{Mg}{I} is significantly greater than that of the other species, due to the fact that the cross-correlation is dominated by the line triplet near 517 nm.}
\label{table:detections}
\centering
\begin{tabular}{l | c c c c c c}
\hline
\hline
Species & A ($\times 10^{-3}$)& $v_0$ (km/s) & FWHM (km/s) & Model (\#2) discrepancy \\
\hline
\ion{Mg}{I} & $1.23  \pm 0.15$ &  $36.36 \pm 0.75$ & $12.4   \pm 1.8$ & $6.97 \pm 0.87$\\
\ion{Ca}{I} & $0.37  \pm 0.056$ & $37.6  \pm 1.1$  & $14.8.  \pm 2.6$ & $2.17 \pm 0.33$\\
\ion{V}{I}  & $0.247 \pm 0.049$ & $35.7  \pm 1.1$  & $11.0   \pm 2.5$ & $1.58 \pm 0.32$\\
\ion{Cr}{I} & $0.224 \pm 0.029$ & $36.40 \pm 0.88$ & $14.0   \pm 2.1$ & $3.23 \pm 0.42$\\
\ion{Fe}{I} & $0.393 \pm 0.024$ & $34.37 \pm 0.48$ & $15.9   \pm 1.1$ & $4.70 \pm 0.29$\\
\ion{Ni}{I} & $0.313 \pm 0.067$ & $37.21 \pm 0.92$ & $8.7    \pm 2.2$ & $8.2  \pm 1.8$\\

\end{tabular}

\end{table*}

\noindent In section \ref{sec:templates} we introduced four models that are injected into the data to enable model comparison. The resulting cross-correlation strengths associated with these is overlaid onto the data in Figures \ref{fig:detections}, \ref{fig:molecules} and \ref{fig:ions}, and the ratio of the injected cross-correlation strength with the observed strength is provided in Table \ref{table:detections}. Given this comparison, we can draw the following conclusions:

\begin{enumerate}
\item Like in other ultra-hot Jupiters, \ion{Fe}{I}, \ion{Mg}{I} and \ion{Cr}{I} are among the strongest absorbing species, with \ion{Mg}{I} absorption (dominated by the resonant triplet near 517 nm) being optically thick up to altitudes significantly greater than the other species. The effect of line absorption by \ion{Fe}{I} and other metals in the high-resolution transmission of WASP-121 b has already been observed by three independent studies by \citet{Gibson2020}, \citet{Cabot2020} and \citet{Bourrier2019}. Our analysis confirms the presence of metal absorption claimed by these authors.

\item Ions are not detected despite the strong detections of \ion{Fe}{II} and \ion{Mg}{II} presented by \citet{Sing2019} in the NUV. This is because these NUV lines are resonant lines, which have much higher oscillator strengths than any lines of these species in the optical. Moreover, due to their smaller oscillator strengths, the absorption lines probed in the optical become optically thick at higher pressures (i.e. lower altitudes) than those in the NUV, where the degree of thermal ionization is higher (see Fig. \ref{fig:chemistry_all}). Interestingly, \ion{Fe}{II} has recently been detected in the high-resolution transmission spectrum of MASCARA-2 b \citep{Stangret2020,Hoeijmakers2020}, which has an equilibrium temperature that is approximately 100 K cooler than WASP-121 b. A detection of \ion{Fe}{II} was recently reported in \citet{Ben-Yami2020}, who use the same data as the present study. This \ion{Fe}{II} feature appears to extend over a wide range of values for the assumed orbital velocity at a relatively low significance, and has an average line-depth of 0.23\%, significantly greater than the noise level apparent in our cross-correlation function (see Fig. \ref{fig:ions}). We therefore conclude that this is likely a false-positive detection. Although discrepancies in the results of cross-correlation based studies are sensitive to the choice of template, we note that the \ion{Fe}{II} template used in this study has been applied successfully to detect the \ion{Fe}{II} feature in MASCARA-2 b, which has a temperature that is similar to WASP-121 b \citep{Hoeijmakers2020}.

\item The absorption strengths of all atoms are under-predicted by factors between 1.5 to 8. We hypothesise that this is due to the atmosphere being extended beyond what is expected assuming hydrostatic equilibrium. The presence of an extended atmosphere has already been evidenced by observations of ionised metals in the near UV \citep{Sing2019} and strong H-$\alpha$ absorption in the optical \citep{Cabot2020}. Significant underprediction by hydrostatic models is common in earlier analyses of other ultra-hot Jupiters, e.g. \citet{Hoeijmakers2018,Hoeijmakers2019}. We note that most of the detections presented in this work would not have been possible without the presence of an extended atmosphere, and that the planning of spectroscopic observations of WASP-121 b or similar planets should take this enhancement of atomic spectral lines into account. We therefore advise against the use of model-dependent metrics to predict which atomic species might be detectable using cross-correlation spectroscopy. For example, the metric introduced by \citet{Ben-Yami2020} predicts that \ion{Ca}{I} and \ion{Mg}{I} should be difficult to observe and these species were consequently excluded from their analysis, even though these are confidently detected in the present study using the same data.

\item  \ion{Ti}{I} is not detected, and indeed is not expected to be a stronger absorber than \ion{V}{I} according to our model. This is discussed further in the next section.

\item This analysis is not expected to be sensitive to VO, H$_2$O nor SH, even though these have significant optical absorption bands. This is primarily due to masking by other lines, which is notably relevant for SH at blue optical wavelengths and TiO/VO, that compete with each other across most of the optical. H$_2$O is expected to be masked by TiO/VO at all wavelengths. Secondly, lines within molecular bands are closely separated, and form blends when taking into account broadening caused by the finite spectral resolution and the rotation of the planet, which dominates at $v_{\textrm{eq}} = 7.0$ \kms. Finally, although the four molecules in question have rich absorption bands with a relatively large integrated opacity, the opacity of individual lines is small compared to the opacity of single atomic lines. This makes the depths of molecular bands weaker than the depths of atomic lines. All three of these effects are visually evident from Fig. \ref{fig:models} and \ref{fig:templates}. The consequences for the SH molecule are discussed further in Section \ref{sec:SH}.
\item TiO could have been detected at an expected abundance of $\sim 2\times10^{-6}$ (see Fig. \ref{fig:molecules}), assuming a reliable line list is used. \citet{Merritt2020} presents a deep search for TiO and VO in similar transit observations of WASP-121 b using the UVES spectrograph, and find an upper limit on the TiO abundance of $10^{-9.26}$, far below  the TiO abundance expected from our model. This limit is significantly deeper than the sensitivity of these data, primarily due to the fact that UVES has additional coverage of red wavelengths where TiO has strong absorption bands. \citet{Merritt2020} also report an upper limit of $10^{-7.88}$ on the abundance of VO, but note that the most accurate line list presently available for VO \citep[by the Exomol group ][]{Tennyson2012} is not accurate enough for application of the cross-correlation technique.
\end{enumerate}

\subsection{Ti-V chemistry and condensation}\label{sec:TiV}
This analysis provides direct evidence for the existence of neutral, atomic vanadium in the atmosphere of WASP-121 b. A detection of \ion{V}{I} contrasts with a non-detection of \ion{Ti}{I}, although \ion{Ti}{I} is approximately 10$\times$ more abundant than \ion{V}{I} in the Sun and hence could be expected to be a stronger absorber, as is the case in stars \citep{Asplund2009}. This section aims to elucidate this dichotomy.

\bigskip

\noindent Fig. \ref{fig:chemistry_all} shows the abundance profiles resulting from our equilibrium chemical model at 2,000 K, $20\times$ solar metallicity. Under these conditions, \ion{Ti}{} and \ion{V}{} are mostly locked up in their respective oxides, with TiO being an order of magnitude more abundant than VO. However, the transmission spectrum observed by \citet{Evans2018} shows the presence of VO bands, while TiO bands appear to be absent. \citet{Evans2018} suggest that this indicates that TiO is condensed out of the atmosphere, due to it having a higher condensation temperature than VO \citep{Lodders2002}. In this scenario, \ion{Ti}{I} would be depleted along with TiO, as chemical equilibrium drives the remaining atomic Ti into its oxide form in response to the decreasing TiO abundance until all Ti bearing species are condensed out of the gas phase.\\

\noindent Conversely, a detection of \ion{V}{I} implies that vanadium is not depleted, and therefore a significant amount of vanadium should exist in the form of gaseous VO. Indeed, our equilibrium-chemistry calculation indicates that between the mbar and nano-bar levels where transmission spectroscopy is sensitive, the VMR of VO is expected to be greater than $\sim 5\times10^{-7}$ (see Fig. \ref{fig:chemistry_all}). The current findings therefore support the interpretation by \citet{Evans2018} that the banded structure in the optical transmission spectrum is caused by VO absorption. This conclusion contrasts with the interpretation in \citet{Ben-Yami2020}. These authors proposed that a detection of \ion{V}{I} is evidence of dissociation of VO that can explain the non-detection of VO by \citet{Merritt2020}, who infer an upper limit on the VO VMR of $10^{-7.9}$, which would amount to a depletion compared to a $20 \times$ solar metallicity gas at 2,000 K in chemical equilibrium. Instead, we argue that if the atmosphere is indeed close to chemical equilibrium, this detection of \ion{V}{I} is evidence for the presence of VO, not its depletion, and we agree with \citet{Merritt2020} that a non-detection of VO using high resolution optical spectroscopy cannot be confidently interpreted while line-list inaccuracies persist.\\

At face value, our non-detection of \ion{Ti}{I} is consistent with the interpretation of TiO depletion put forth by \citet{Evans2018}, and this position is adopted by \citet{Ben-Yami2020} as well. However, we observe that the injected model spectra (see Fig. \ref{fig:models}) predict that \ion{Ti}{I} is a weaker absorber than \ion{V}{I}, even in chemical equilibrium, and is hence not expected to be detectable despite the fact that \ion{Ti}{} is approximately $10\times$ more abundant than \ion{V}{} in a gas with solar elemental abundance ratios \citep{Asplund2009}.\\

\begin{figure*}
    \includegraphics[width=\textwidth]{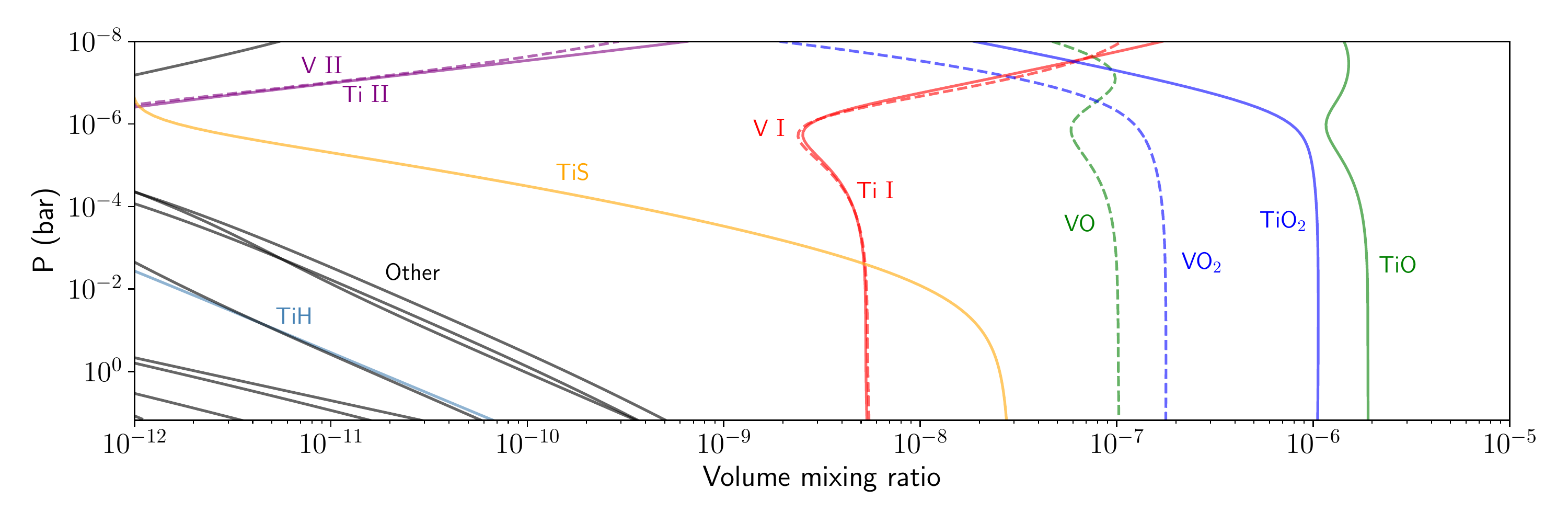}
	\caption{Abundance profiles of all Ti and V-bearing species included in \texttt{FastChem}, at a temperature of 2,000 K and a metallicy of $20\times$ solar. All solid lines correspond to Ti-bearing species, while dashed lines correspond to V-bearing species. This figure supports the conclusions that \ion{Ti}{} and \ion{V}{} are expected to have equal abundance; that both mono-oxides as well as  dioxides are important constituents of the atmosphere, as well as metal sulphites; and that \ion{Ti}{II} and \ion{V}{II} are not important in chemical equilibrium.}
	\label{fig:TiV}
\end{figure*}

\noindent To explain this, we investigate the abundances of all Ti and V-bearing species presently included in \texttt{FastChem}, which are shown in Fig. \ref{fig:TiV}. Our model indicates that for a $20\times$ solar metallicity gas at 2,000 K in chemical equilibrium, \ion{Ti}{I} is more efficiently converted into molecules than \ion{V}{I}, to the point that the abundances of \ion{Ti}{I} and \ion{V}{I} are nearly equal (around $5\times 10^{-9}$, see the red curves in Fig. \ref{fig:TiV}). Although \texttt{FastChem} includes more Ti-bearing species than V-bearing species, over most of the atmosphere these additional Ti-bearing molecules do not represent a significant fraction of the \ion{Ti}{I} reservoir, and can therefore not explain the observed disappearance of the abundance difference between \ion{Ti}{I} and \ion{V}{I}. The exception is TiS, which is more abundant than \ion{Ti}{I} below altitudes at a pressure of $\sim 1$ mbar. Presumably, the VS molecule could have a similarly significant abundance, but it is not included in \texttt{FastChem}, which could lead \texttt{FastChem} to overestimate the \ion{V}{I} abundance. However, we note that the abundance profile of \ion{Ti}{I} does not seem to react to the steeply decreasing abundance of TiS with altitude. This leads us to conclude that the claim that \ion{Ti}{I} and \ion{V}{I} have equal abundance is robust against the fact that VS is missing from \texttt{FastChem}.\\

\noindent It is apparent from Fig. \ref{fig:TiV} that the dioxide molecules TiO$_2$ and VO$_2$ represent important fractions of the total Ti and V reservoirs. Especially VO$_2$ is expected to be approximately twice as abundant as VO. As noted previously, we observe that TiS is an important constituent as well, but that TiH is not, with a diminishing abundance that does not exceed $10^{-10}$. Although we presently have no knowledge of the absorption cross-sections of gaseous TiO$_2$ and VO$_2$ at high temperatures, our model indicates that these molecules can be important sources of opacity if they are optically active to a similar degree as their mono-oxide counterparts.
TiO$_2$ is known to exhibit electronic transitions at optical wavelengths at cryogenic temperatures \citep{Garkushka2008}. TiS bands have been observed in the spectra of Mira-type AGB stars between 800 nm and 1.35 \um \citep{Jonsson1992}, and VS is measured to be active at near-infrared wavelengths \citep{Bauschlicher1986}. We note that a significant body of experimental observations of the TiS molecules exists in the literature, \citep[e.g.][]{Jonsson1993,Ran1999,Cheung2000,Pulliam2010}, which may make TiS amenable for detection in future cross-correlation studies.\\

\noindent We expect that spectral retrievals performed on HST/STIS and WFC3 observations of transmission and emission spectroscopy of hot Jupiters \citep[e.g.][a.o.]{Evans2018} may yield biased measurements of the chemical composition if significant opacity sources are not taken into account. We therefore emphasise the importance of the investigation of the absorption cross-sections of more complex metal-bearing molecules.\\

\noindent Finally, we note that although chromium and nickel condensation chemistry has not yet been extensively explored in the exoplanet literature, the present observations suggest condensation of these species commences at lower temperatures than titanium. This supports the notion that \ion{Ti}{} is one of the first species to condense out of the atmosphere, with a condensation temperature roughly 200 K higher than \ion{Cr}{} and \ion{Ni}{} \citep{Lodders2002,Lodders2003}.\\

\subsection{The NUV slope and absorption by SH}\label{sec:SH}
\citet{Evans2018} reported the presence of a strong increase of the transit radius towards bluer wavelengths. Such slopes are commonly observed in the transmission spectra of hot Jupiters \citep[e.g.][]{Sing2016}, and are primarily associated with Rayleigh scattering of H$_2$ or aerosols \citep{Lecavelier2008,Lecavelier2008a,Pont2008}. When caused by H$_2$ scattering, the slope can be used to measure the atmospheric temperature because the underlying opacity function is known. This scenario is ruled out by \citet{Evans2018}, because the strength of the slope in the transmission spectrum of WASP-121 b would require a temperature of $6980\pm3660$ if caused by pure H$_2$ scattering. This temperature is much higher than the measured day side temperature \citep{Bourrier2019}, and thermal dissociation would preclude the existence of significant amounts of H$_2$ \citep{Kitzmann2018}. \citet{Evans2018} therefore hypothesise that this NUV absorption is caused by additional molecular absorption bands, and propose the SH molecule as an explanation because it has strong electronic absorption bands shortwards of $\sim 450$ nm and can exist at significant abundances \citep{Zahnle2009}.\\

\noindent We included SH opacity in the model spectra used in the this work, assuming an ad hoc (non-equilibrium) abundance of $10^{-4}$ for pressures higher than $10^{-5}$ bar, approximately matching the profile assumed by \citet{Evans2018}. The SH line-list recently published by the exomol group \citep{Gorman2019} is expected to be calculated at sufficient accuracy for the application of the cross-correlation technique (Yurchenko \& Tennyson, private communication). However regardless of line list accuracy, our models indicate that even at an abundance of $10^{-4}$, the absorption bands by SH are significantly weaker than typical lines by atomic metals, causing the SH lines to be masked as already noted by \citet{Lothringer2020}. The effect of masking is evident from both Figures \ref{fig:templates} and \ref{fig:models}, which show that the increase in transit radius in the SH bands is small compared to those of atomic metal lines, which increase in strength and number towards shorter wavelengths. The result of this masking is also evident from the model comparison after cross-correlation, shown in Fig. \ref{fig:molecules}. Despite the presence of a plenitude of absorption lines in the HARPS bandpass between 373 and $\sim 450$ nm, the application of the cross-correlation analysis to the model-injected spectra does not result in an expected detection unless the model is free from absorbers other than SH. \\

\begin{figure}
    \includegraphics[width=0.5\textwidth]{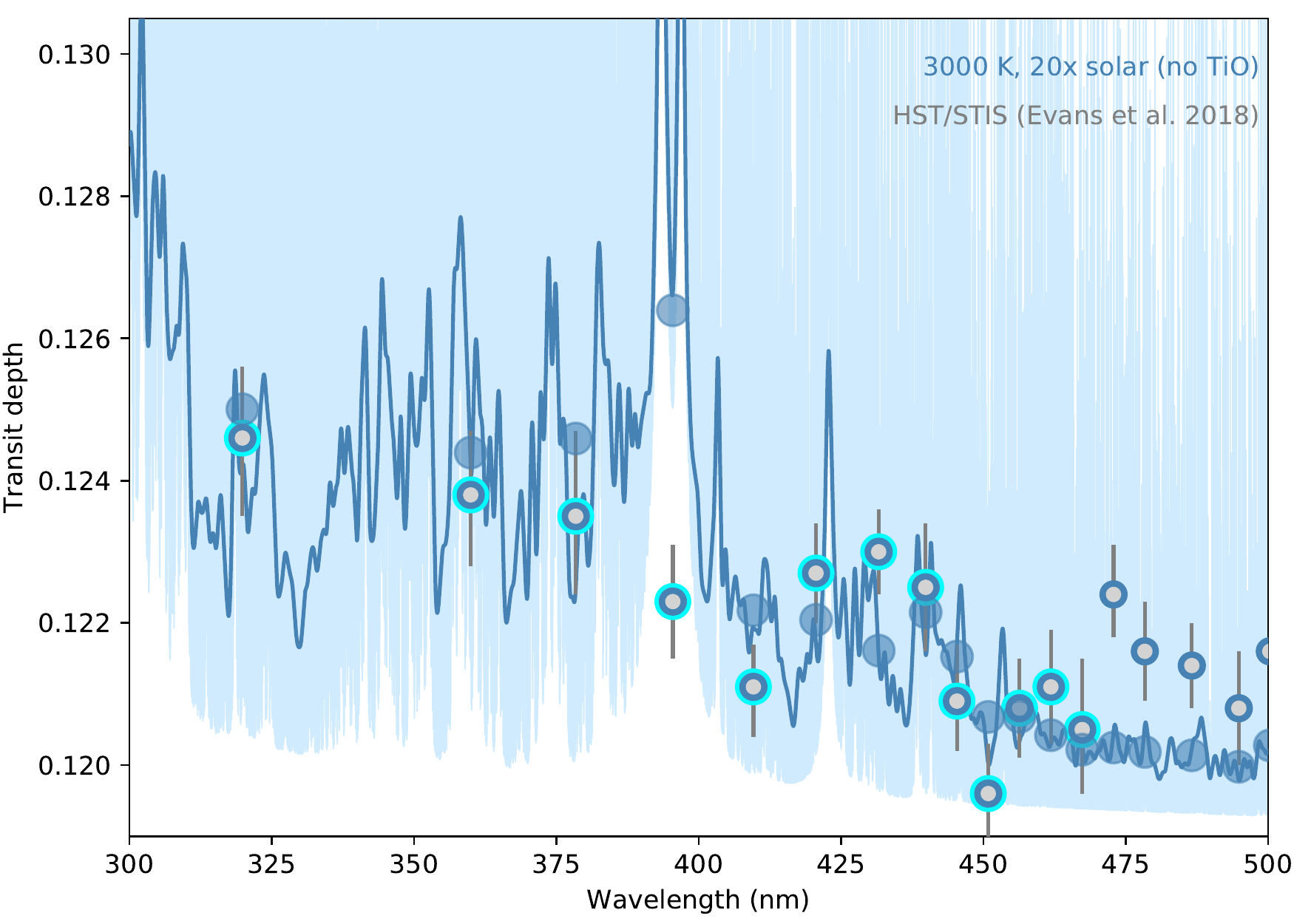}
	\caption{STIS/G430L transmission spectrum of WASP-121 b as published by \citet{Evans2018} indicated with open circles. A model spectrum assuming chemical equilibrium, a temperature of 3,000 K, 20$\times$ solar metallicity and SH computed as described in Section \ref{sec:templates} and Table \ref{table:modelsoverview} is overplotted, both at the native resolution (light-blue) and binned to the resolution of STIS (dark-blue line and circles). The wavelength bin at 395 nm coincides with the \ion{Al}{I} doublet, at which location the transit radius is strongly over-predicted compared to the STIS observations. This models predicts that the cores of the \ion{Al}{I} doublet become optically thick at transit radii of $\sim 5\times10^{-3}$ above the local continuum, which corresponds to an altitude of $\sim 0.3 R_p$ above the planet. At such low pressures, photo-ionization is expected to be important \citep[e.g.][]{Moses2011,Kitzmann2018}. At the same time, \ion{Al}{I} has one of the lowest ionization thresholds of the elements, so we predict that equilibrium models may strongly over-predict the strength of the \ion{Al}{I} doublet at this wavelength.}
	\label{fig:STIS}
\end{figure}

\noindent Strong absorption by metals with large NUV cross-sections has already been observed as part of previous studies \citep{Sing2019,Gibson2020,Cabot2020}, as well as the present work. These detections increase likelihood that the observed NUV opacity can be explained by atomic metal lines alone. Fig. \ref{fig:STIS} shows the transmission spectrum of WASP-121 b as observed by \citet{Evans2018} using the STIS instrument, together with a model that assumes equilibrium chemistry, $20\times$ solar metallicity and an isothermal profile of 3,000 K, binned to the resolution of the STIS spectrum. This model shows that if the temperature of the atmosphere of WASP-121 b is elevated (e.g. in the case of a significant thermal inversion), atomic metals alone can account for the observed NUV slope without breaching chemical or hydrostatic equilibrium \citep{Lothringer2020}. Moreover, we note that the anomalously strong metal lines detected by \citet{Sing2019} and \citet{Gibson2020} as well as this study, indicate that the atmosphere is significantly inflated, further strengthening the above hypothesis.\\

\subsection{Radial velocity offset and winds}
The cross-correlation with the \ion{Fe}{I} template can be used as a probe to measure the radial velocity of the star. We select the out-of-transit spectra taken in each of the three nights, and fit the centroid position of the average stellar \ion{Fe}{I} line within 15 \kms of the core - approximately equal to the projected rotational velocity. In this way, we obtain measurements of the systemic radial velocity of $38.041 \pm 0.0044$, $38.0393 \pm 0.0028$ and $38.0487 \pm 0.0034$ \kms for each of the three nights respectively, or $38.0430 \pm 0.0021$ \kms on average. \\

\noindent The absorption lines of the planet appear at systemic velocities of 34 to 37 \kms, indicative of significant blue shifts of the absorbing species in the planet atmosphere. The signature of \ion{Fe}{I} appears most shifted, with an effective shift of $38.0430 \pm 0.0021 - 34.37 \pm 0.48 = 3.67 \pm 0.48$ \kms, consistent with values found in the recent literature, of $4.3 \pm 0.56$ \kms and $5.2 \pm 0.5$ \kms by \citet{Gibson2020} and \citet{Bourrier2020} respectively. The observation that different atoms exhibit significantly different blue-shifts may be evidence that the atmosphere is dynamically heterogeneous, or that species are distributed heterogeneously around the limb in a fashion similar to what has been observed on WASP-76 b \citep{Ehrenreich2020}.

\section{Conclusion}
This paper presents an analysis of the optical transmission spectrum of WASP-121 b using spectroscopic observations obtained with the HARPS spectrograph during three transit events. WASP-121 b is an ultra-hot Jupiter in the orbit of a bright (V=10.4) F6V star with a period of 1.27 days. The planet has been targeted by a series of observational campaigns using the WFC3 and STIS spectrographs aboard the Hubble Space Telescope \citep{Evans2016,Evans2017,Evans2018,Evans2019,Sing2019}, which revealed a rich transmission spectrum with signatures of absorption bands of VO, while TiO appears depleted because of condensation. The NUV spectrum shows a strong slope, which was hypothesised to be caused by the SH radical, which may exist as a photo-chemical by-product. In addition, deep ionic UV lines  of \ion{Fe}{II} and \ion{Mg}{II} as well as H-$\alpha$ are evidence the existence of an extended, possibly escaping envelope \citep{Sing2019,Cabot2020}.\\

\noindent In this work we search the transmission spectrum for line absorption of the \ion{Na}{I} doublet and perform a survey of additional atomic species with rich absorption spectra, as well as TiO, VO, H$_2$O and SH. We compare the resulting cross-correlation functions with forward models of the transmission spectrum of WASP-121 b assuming chemical equilibrium \citep[computed with \texttt{FastChem},][]{Stock2018} at temperatures of 1,500 K, 2,000 K and 3,000 K, and elemental abundance ratios set to $20\times$ the solar metallicity; apart from SH which is a non-equilibrium species. The findings of this work are summarised as follows:

\begin{itemize}
    \item The \ion{Na}{I} doublet is detected in the transmission spectrum at a confidence of $8.6\sigma$, with an average absorption depth of $(5.59 \pm 0.65) \times 10^{-3}$ times the flux of the host star (see Section \ref{sec:sodiumresults}, and Fig. \ref{fig:NaSpectrum}).
    \item The \ion{Na}{I} lines are significantly broadened compared to what is expected from hydrostatic models and rigid-body rotation of the planet when assuming tidal locking, and the line depths show a difference corresponding to $15.5\pm5.1$ times the atmospheric scale height (assuming a temperature of 2,000 K and a mean particle weight of 2.3 u). This could indicate that \ion{Na}{I} forms an optically thin envelope around the planet \citep{Oza2019}. An analytical description of a homogeneous optically thin cloud with an integrated sodium mass of $(3.6 \pm 0.6) \times10^{10}$ g reproduces the transmission spectrum well, as does a model by \citet{Wyttenbach2020} of a hydrodynamic outflow, and a model by \citet{Gebek2020} that assumes an optically thin toroidal \ion{Na}{I} envelope (see Section \ref{sec:sodiumresults}, Fig. \ref{fig:NaSpectrum} and Appendix \ref{app:sodium}).
    \item Via application of the cross-correlation technique we detect neutral magnesium, calcium, vanadium, chromium, iron and nickel (see Section \ref{sec:detections}). The line-strengths of all detected species are under-predicted by model spectra that assume chemical equilibrium, which strengthens existing evidence that the atmosphere is not in hydrostatic equilibrium. In addition, the lines of all detected species are blue shifted by 1 to 4 \kms compared to the planet rest-frame, which is evidence of atmospheric dynamics (see Table \ref{table:detections}), and consistent with previous findings \citep{Bourrier2020,Cabot2020,Gibson2020}.
    \item The presence of \ion{V}{I} implies the presence of VO, which is expected to exist in chemical equilibrium at much higher abundance than \ion{V}{I}. These observations therefore support the interpretation by \citet{Evans2018} that molecular bands in the transmission spectrum are caused by VO (see Section \ref{sec:TiV} and Fig. \ref{fig:TiV}), instead of being indicative of VO depletion \citep{Ben-Yami2020}. We emphasise that careful model comparison is crucial to interpret the various detections and non-detections of atomic species and their chemical interactions with molecules (see below).
    \item Ions, including \ion{Fe}{II} are searched for using hot (4,000 K) templates earlier applied to observations of KELT-9 b \citep{Hoeijmakers2018,Hoeijmakers2019}, but are not detected. Ions are not expected to be important atmospheric constituents in chemical equilibrium (see Figures \ref{fig:ions} and \ref{fig:TiV}). Although strong \ion{Fe}{II} lines have recently been observed at high altitudes $\sim 2 R_p$ using HST/STIS in the UV by \citet{Sing2019}, these lines are strong resonant lines with higher intrinsic opacities than lines in the optical. Having weaker oscillator strengths, these lines are formed at higher pressures than the resonant lines in the UV, where the temperature and degree of ionization may be significantly lower. We interpret a recent detection of \ion{Fe}{II} in the optical transmission spectrum of this planet \citep{Ben-Yami2020} as a likely false-positive, though further observations and comparative analyses may be needed to confirm this.
    \item Our models indicate that chemical equilibrium drives the abundance of \ion{Ti}{I} and \ion{V}{I} to near-equal values despite the fact that \ion{Ti}{I} is $\sim 10\times$ more abundant in the Sun. This makes \ion{V}{I} a stronger absorber than \ion{Ti}{I} over this wavelength range (see Section \ref{sec:TiV} and Fig. \ref{fig:TiV}) and implies that a non-detection of \ion{Ti}{I} alone in this dataset is not sufficient to conclude that \ion{Ti}{} is depleted from the atmosphere as suggested in \citet{Ben-Yami2020}.
    \item Generally, molecules are difficult to detect due to masking, line broadening and the fact that bands intrinsically have shallower depths (i.e. effective transit radii) than individual atomic lines. No signatures of TiO, VO, H$_2$O or SH are detected (see Section \ref{fig:molecules}).
    \item Moreover, atoms would not or barely be observable if the atmosphere at the terminator was in hydrostatic equilibrium at 2,000 K (see Fig. \ref{fig:detections}), unless photo-dissociation and vertical mixing are capable of significantly increasing the abundances atomic species at higher altitudes (see Fig. \ref{fig:detections}). If the temperature in the upper atmosphere is significantly higher than 2,000 K, this could act to increase the scale height in hydrostatic equilibrium \citep[e.g.][]{Wyttenbach2017}. However, this would increase the ionization fraction, in turn reducing opacity by neutral species.
    \item Without condensation, TiO is expected to reach an abundance of $2\times 10^{-6}$ in chemical equilibrium, which we expect to be barely detectable with the current data (see Fig. \ref{fig:molecules}). Our non-detection of TiO is consistent with a deeper upper limit recently established by \citet{Merritt2020}, and both observations are consistent with the hypothesis by \citet{Evans2018} that TiO is subject to condensation and rain-out.
    \item Titanium and vanadium dioxides and sulphides are expected to be significant atmospheric constituents at 2,000 K (see section \ref{sec:TiV} and Fig. \ref{fig:TiV}) and may cause significant absorption at optical or near-infra-red wavelengths \citep{Bauschlicher1986,Jonsson1992,Garkushka2008}. Currently, accurate line-lists do not exist for these molecules, impeding the usage of the cross-correlation technique to detect them in future studies.
    \item Given that atomic metal lines are inflated beyond what is predicted by models in hydrostatic equilibrium at 2,000 K, it is plausible that absorption by atomic metals alone is sufficient to explain the strong NUV slope detected with by \citet{Evans2018} with HST/STIS (see Section \ref{sec:SH} and Fig. \ref{fig:STIS}).
\end{itemize}

\begin{acknowledgements}
This work has been carried out in the framework of the PlanetS National Centre of Competence in Research (NCCR) supported by the Swiss National Science Foundation (SNSF). It has received funding from the European Research Council (ERC) under the European Union’s Horizon 2020 research and innovation programme (projects Four Aces, EXOKLEIN and Exo-Atmos with grant agreement numbers 724427, 771620 and 679633, respectively), and STFC Project No. ST/R000476/1. A.W. acknowledges the financial support of the SNSF by grant number P400P2\_186765. N. A.-D. acknowledges the support of FONDECYT project 3180063.
The analysis presented in this work has made use of the VALD database, operated at Uppsala University, the Institute of Astronomy RAS in Moscow, and the University of Vienna; Ian Crossfields' Astro-Python Code library and Astropy,\footnote{\url{http://www.astropy.org}} a community-developed core Python package for Astronomy \citep{astropy:2013, astropy:2018}. We thank our anonymous referee for his/her thorough reading and constructive commentary, which helped us to significantly improve this manuscript and refine our understanding.
\end{acknowledgements}

\bibliographystyle{aa}
\bibliography{main_arxiv}

\begin{appendix}
\section{Sodium lines in the optically thin limit}\label{app:sodium}
Following the argument expanded by \citet{Oza2019} and \citet{Gebek2020}, we start with the flux emanating from a source with spectrum $F_0(\lambda)$ that is blocked by a cloud of with optical depth $\tau(\lambda)$ located in the line of sight to the source \citep{Draine2011}:
\begin{equation}
    F(\lambda) = F_0(\lambda) e^{-\tau(\lambda)}
\end{equation}

\begin{figure}[h!]
    \centering
    \includegraphics[width=0.94\columnwidth]{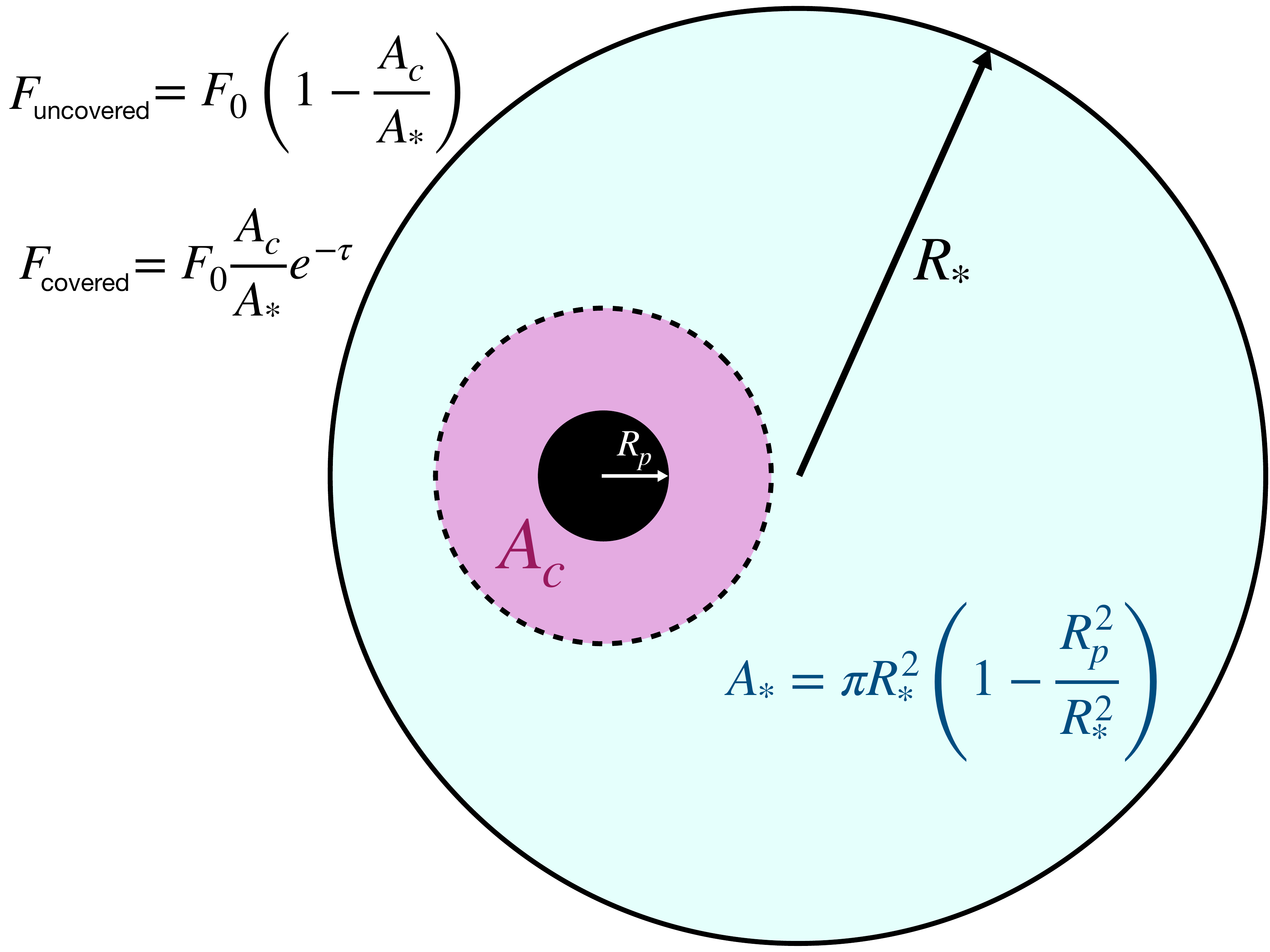}
	\caption{Schematic of the geometry of an optically thin cloud with projected area $A_c$ surrounding an optically thick planet with radius $R_p$, both transiting a star with radius $R_*$, with an effective projected area $A_*$. Equation \ref{eq:flux} consists of the sum of the covered and uncovered components, $F_{\textrm{covered}}$ and $F_{\textrm{uncovered}}$ respectively.}
	\label{fig:schematic}
\end{figure}

\noindent This is valid if the source is totally covered by the foreground cloud. When considering a cloud with projected area $A_c$ and optical depth $\tau$ in front of a larger star with projected area $A_*$ (see Fig. \ref{fig:schematic}), this equation modifies to:

\begin{equation}\label{eq:flux}
    F(\lambda) = F_0(\lambda)\frac{A_c}{A_*}e^{-\tau(\lambda)}+F_0(\lambda)\left(1-\frac{A_c}{A_*}\right).
\end{equation}

\noindent Rewriting, the line-depth with respect to the continuum is:

\begin{equation}
    F_0 - F = F_0\frac{A_c}{A_*}\left(1-e^{-\tau}\right),
\end{equation}

or in units of \textit{relative transit depth}, which corresponds to the equivalent width as defined by \citep{Heng2015}:

\begin{equation}\label{eq:linedepth}
    \frac{F_0 - F}{F_0} \equiv W = \frac{A_c}{A_*}\left(1-e^{-\tau}\right) = \frac{A_c}{A_*} \tau,
\end{equation}

in the optically thin limit, i.e. $\tau \ll 1$.  The optical depth is either expressed in terms of a cross section multiplied by a column density (with units of cm$^{-2}$),

\begin{equation}
\tau = \sigma N,
\end{equation}

or an opacity multiplied by a mass column density (with units of g cm$^{-2}$),

\begin{equation}
\tau = \kappa \tilde{M}.
\end{equation}

\noindent For a line transition with oscillator strength $f$, we have:

\begin{equation}
\sigma = \frac{\pi e^2 f}{m_e c} \Phi, \textrm{~and}~\kappa = \frac{\sigma}{m_{\rm Na}}.
\end{equation}

\noindent Consider a Gaussian line profile $\Phi$, which is broadened due to some non-thermal velocity $\Delta v$.  We assume that $\Delta v$ is a constant fitting parameter:
\begin{equation}
\Phi = \frac{c}{\nu_0}\frac{1}{\Delta v \sqrt{2\pi}} ~e^{-\left(\frac{c (\nu_0-\nu)}{\sqrt{2} \Delta v \nu_0}\right)^2},
\end{equation}

\noindent The coefficient of $\Phi$ is set by the requirement that $\int \Phi ~d\nu=1$.
Rewriting, we obtain the optical depth at the line-center \citep{Draine2011}:

\begin{equation}
    \tau_0 = \frac{\sqrt{\pi} e^2}{m_e c} \frac{N f \lambda_0}{\sqrt{2}\Delta v}.
\end{equation}

\noindent Or, in terms of mass column density and the transit depth:

\begin{equation}
    W(\lambda_0) = \frac{A_c}{A_*} \tau_0 = \frac{A_c}{A_*} \frac{\sqrt{\pi} e^2}{m_e c} \frac{~\tilde{M}}{m_\textrm{Na}}\frac{f \lambda_0}{\sqrt{2}\Delta v}
\end{equation}

\noindent This equation contains four unknowns:  $A_c$, $W$, $\Delta v$ and $\tilde{M}$. Already having fit the line-depth and width of the Gaussian line-profile from the observed transmission spectrum \ref{sec:sodiumresults}), we solve for $\tilde{M}$:

\begin{equation}
    ~\tilde{M} = \frac{A_*}{A_c}\frac{m_e c}{\sqrt{\pi} e^2}\frac{\sqrt{2} \Delta v}{f \lambda_0} m_\textrm{Na} W(\lambda_0).
\end{equation}

\noindent Finally, in order to obtain the total mass of sodium in the cloud, we assume that all sodium is in the ground state \citep[essentially ignoring the effect of a non-LTE level population, see ][]{Fisher2019} and multiply the mass column density with the projected area $A_c$ of the cloud, which subsequently cancels:

\begin{equation}\label{eq:massdensity}
    M_{\rm Na} = A_c ~\tilde{M} = A_*\frac{m_e c}{\sqrt{\pi} e^2}\frac{\sqrt{2} \Delta v}{f \lambda_0} m_\textrm{Na} W(\lambda_0).
\end{equation}

\noindent Finally, filling in the projected area of the star taking into account the presence of the optically thick planet with radius $R_p^2$ (see Fig. \ref{fig:schematic}), i.e.,
\begin{equation}
    A_* = \pi R_*^2 \left( 1 - \frac{R_p^2}{R_*^2} \right).
\end{equation}

we obtain

\begin{equation}
    M_{\rm Na} = \pi R_*^2 \left( 1 - \frac{R_p^2}{R_*^2} \right) \frac{m_e c}{\sqrt{\pi} e^2}\frac{\sqrt{2} \Delta v}{f \lambda_0} m_\textrm{Na} W(\lambda_0).
\end{equation}

\noindent In the optically thin limit, the mass of absorbing sodium atoms is independent of the projected area of the cloud as long as the cloud is homogeneous and optically thin. A dependence on $\pi R_*^2$ remains because the same line-depth measured at a larger star requires the presence of more absorbing atoms, and vice versa.

\clearpage
\section{Cross-correlation analysis procedure}\label{app:analysis_steps}
\bigskip
\begin{figure*}
\centering
    \includegraphics[width=0.9\textwidth, trim=0cm 1cm 0cm 2cm, clip]{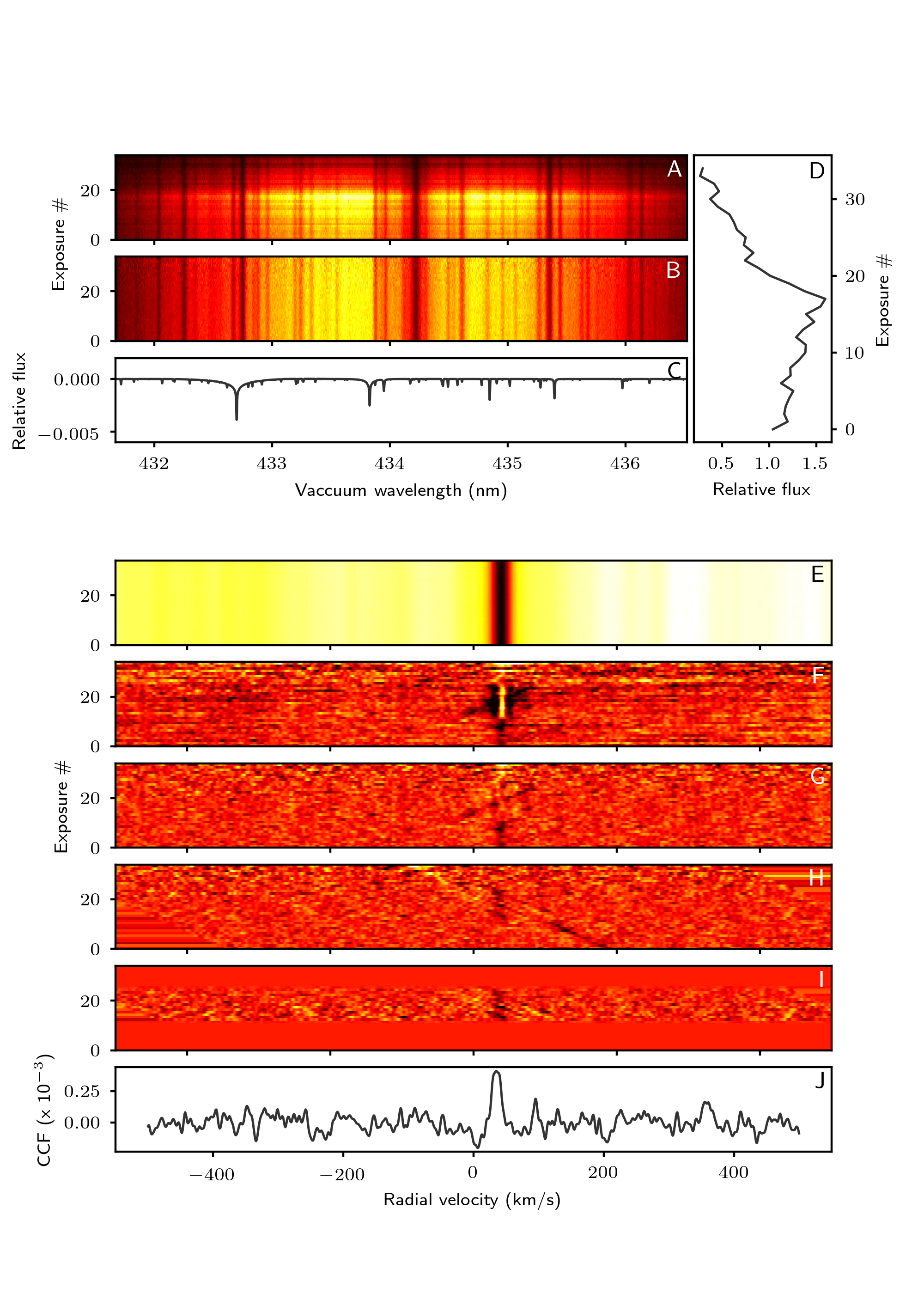}
	\caption{Pre-processing steps applied to obtain cross-correlation functions in the rest-frame of the planet, as described in Section \ref{sec:cross_correlation}. \textbf{Panel A:} An example spectral order of night one as extracted by the HARPS DRS, plotted as a time-series. \textbf{Panel B:} After normalization by the time-average flux and colour correction. \textbf{Panel C:} The cross-correlation template of \ion{Fe}{I} after continuum-subtraction, with which the spectra are cross-correlated. \textbf{Panel D:} The average flux in the spectral order, that is divided out of the spectra in panel A to obtain the spectra in panel B. \textbf{Panel E:} The cross-correlation function of \ion{Fe}{I} of the entire time series. \textbf{Panel F:} After removal of the time-average of the out-of-transit cross-correlation functions. \textbf{Panel G:} After removal of the Doppler shadow and application of the high-pass filter. \textbf{Panel H:} After shifting the cross-correlations to the rest-frame of the planet. \textbf{Panel I} After weighing the cross-correlation functions by the average flux of the time-series (panel D) and selecting only in-transit observations. \textbf{Panel J:} The one dimensional cross-correlation function in the rest-frame of the planet, after averaging the cross-correlation functions (panel I) in time.}
	\label{fig:analysis_steps}
\end{figure*}

\clearpage

\section{Cross-correlation bootstrap analyses}\label{app:bootstrap}
The purpose of the cross-correlation technique is to average hundreds or thousands of absorption lines to reduce the photon-noise and increase detection sensitivity \citep{Snellen2010}. Provided that sufficient spectra and spectral lines available, this technique is known to be capable to reach $1\sigma$ noise levels of $\sim 10^{-6}$ \citep{Hoeijmakers2018b}. There are various effects that may introduce systematic noise structures at this level. Some of these do not correlate with the cross-correlation template (e.g. the presence of uncorrected cosmic-ray hits, uncorrected telluric lines when not correlating with H$_2$O or O$_2$ templates), resulting in pseudo-stochastic variations in the cross-correlation function that can occur at random times or at random radial velocities. Other effects may fully or partially correlate with the cross-correlation template, e.g. residuals from stellar absorption lines (when correlating with templates of species that are present in both the planetary and the stellar spectrum) or aliasing of the cross-correlation template (i.e. when the template has few, or regularly-spaced lines).\\

\noindent These effects cause structures in the cross-correlation function that are not normally distributed, invalidating the usage of summary statistics such as Gaussian confidence intervals to a certain degree. The presence of such noise sources may also cause signatures that mimic planetary absorption lines, leading to the possibility of false-positive detections. We choose to employ two bootstrap methods to investigate the extent at which systematic noise in the cross-correlation function is capable of creating spurious variations that could be mistaken for planetary absorption lines. Both methods are applied to the cross-correlation functions of each species independently, and both methods assume that spurious signals will not be mistaken for planetary absorption lines if they do not resemble Gaussian line profiles with a certain width.

\subsection{Bootstrap method A}
The first method is similar to the bootstrap (EMC) method described in section \ref{sec:dataquality} and \citet{Re08}, in that it constructs distributions using random permutations of in and out-of-transit cross-correlation functions. It aims to assess whether the detected absorption signal is uniquely and evenly present in the in-transit CCFs only, i.e. to reject cases where the observed signal exists in (some of) the out-of-transit CCFs, or is only present in a small fraction of the in-transit CCFs.\\

\noindent We start with the CCFs in which the time-averaged out-of-transit CCF and the best-fit model for the Doppler-shadow are removed. These CCFs of all three nights are split into two groups of in-transit and out-of-transit CCFs. We then obtain `master` in and out-of-transit CCFs by taking the time-average (mean CCF) of both groups. The latter of these should be close to zero, because the time-average out-of-transit CCF had already been removed for each night in the rest-frame of the star. This time-average was a constant, so it cannot have introduced or removed signals in/from only some of the CCFs, as is being tested in this analysis.\\

\begin{figure*}
    \includegraphics[width=0.9\textwidth]{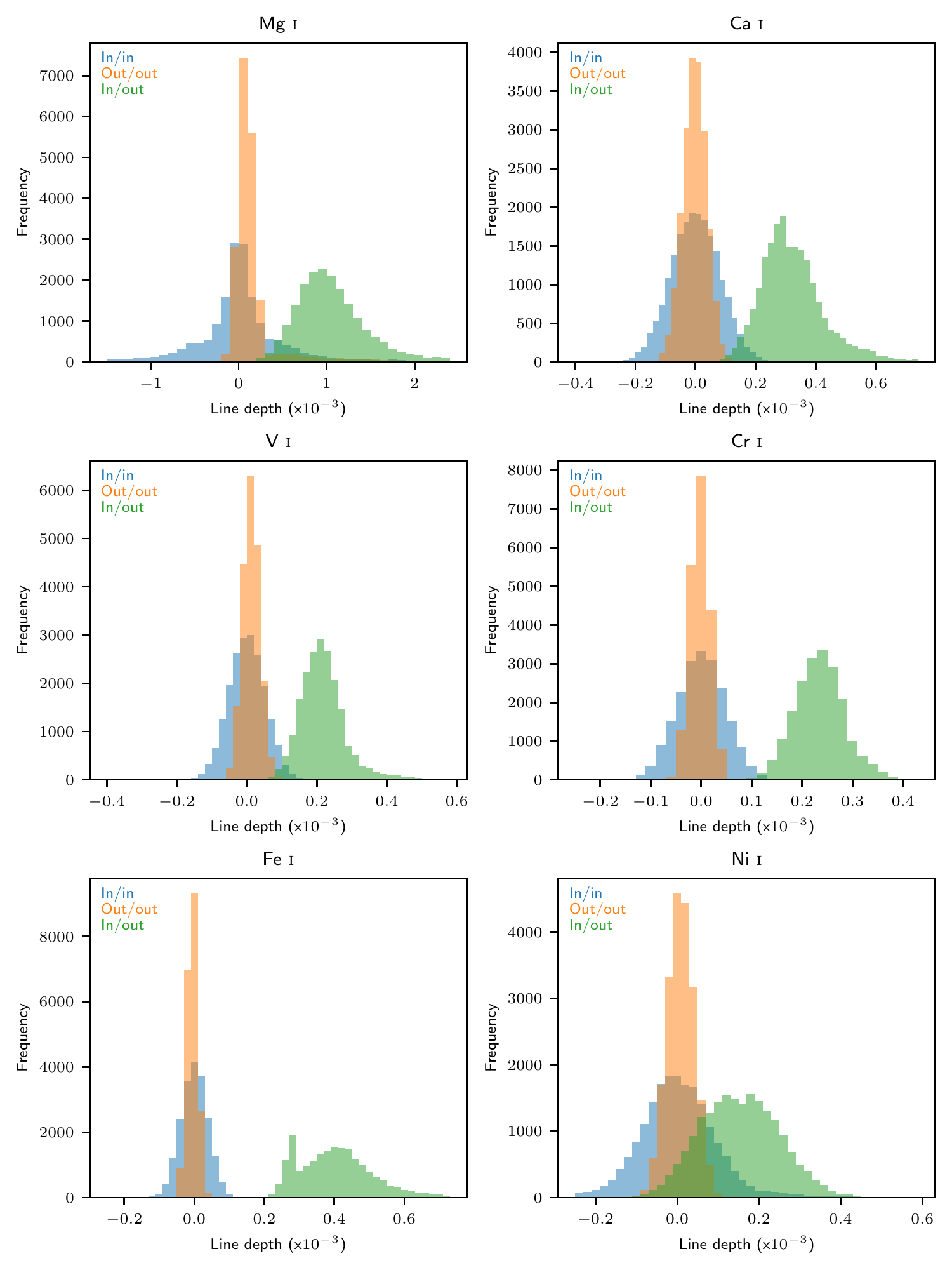}
	\caption{Distributions of randomly generated instances of the 1D CCF at the rest-frame velocity of the planet. The distributions are constructed by considering random subsets of in- or out-of-transit CCFs, for each of the detected species reported in Section \ref{sec:detections}. The detected line amplitude is expected to be zero if out-of-transit CCFs are considered (out-out in orange and if in-transit CCFs are normalized with the master in-transit spectrum (in-in, blue). Only when the in-transit CCFs are normalised by the out-of-transit CCFs (in-out, green) is the measured line amplitude expected to deviate from zero; significantly so for the six species reported.}
	\label{fig:bootstrap_A}
\end{figure*}

\noindent Finally, we randomly select and average 50\% of the in and out-of-transit CCFs, divide these by the master in and out-of-transit CCFs to obtain one-dimensional realizations of in-in, in-out and out-out residuals. To each of these, we fit a Gaussian at the expected location of the planet absorption line within a window of 30 to 44 \kms, corresponding to the observed range of velocities for our detected signals $\pm 4$ \kms, and a width fixed to the measured value (Table \ref{table:detections}). This procedure is repeated 20,000 times and the resulting distributions for each of the detected species are shown in Fig. \ref{fig:bootstrap_A}. These Figures confirm that the detected absorption lines indeed uniquely result from the in-transit spectra, and are not likely to be reproducible via the false-positive scenarios we set up.

\subsection{Bootstrap method B}
The second method aims to assess the distribution of stochastically emerging signals due to variations in the CCF caused by systematic noise that is not correlated with the cross-correlation template. These variations produce structures in the CCF that can occur at any time or radial velocity, independent of the rest-frame velocities of the planet and the star.\\

\noindent Like for method A, we start with the CCFs in which the time-averaged out-of-transit CCF and the best-fit model for the Doppler-shadow are removed, and selecting all CCF values at radial velocities away from the planetary and stellar rest-frames, at times during the planet transit. We select only CCFs obtained during the transit because the statistics of the in-and-out of transit spectra are different. There are more out-of-transit spectra than in-transit spectra, and these tend to be observed at higher airmass. CCF values at velocities occupied by the planet at any instance during the observations (i.e. $\sim \pm 150$ \kms) are ignored, leaving only CCF values that are not directly related to the presence of the signatures of the planet or the star.\\

\noindent Each CCF in the time-series is then shifted to a random radial velocity drawn from a uniform distribution, after which all CCFs are averaged. This process mimicks the co-addition of in-transit CCFs in the rest-frame of the planet, but only evaluated at velocities away from the planet rest-frame and applying an incoherent set of velocity shifts. This method ensures that any systematic correlations that exist purely in the spectral direction of the various spectra of the time-series are preserved. \\

\begin{figure*}
    \includegraphics[width=0.9\textwidth]{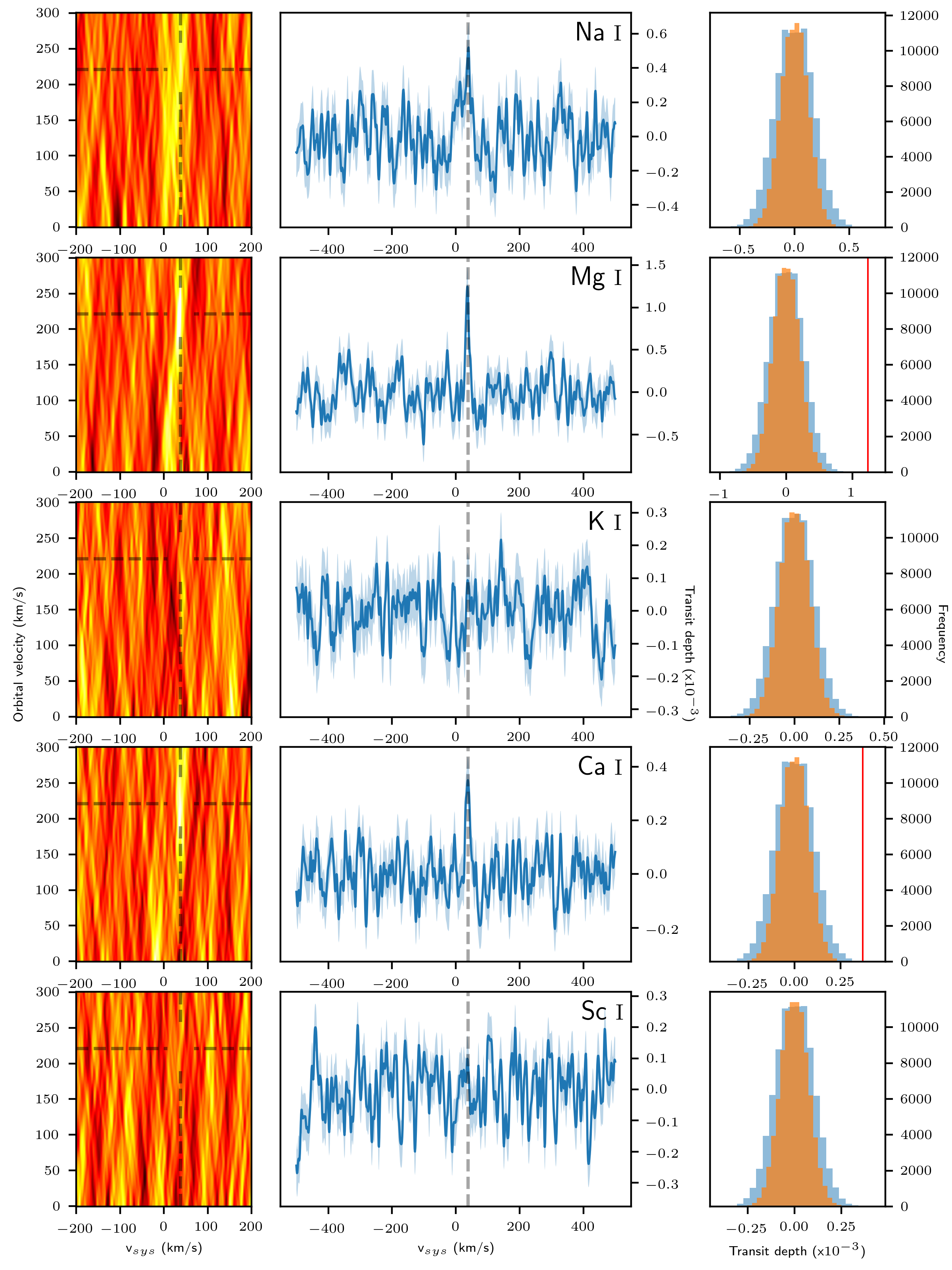}
	\caption{All cross-correlation functions in Kp-Vsys space (left column), and co-added in the rest-frame of the planet (middle column). The blue shaded area indicates the $1-\sigma$ error expected from photon-noise. The right column shows distributions of Gaussian fits to random realisations of the 1D-cross-correlation function generated by stacking the time-series of cross-correlation functions with random shifts applied to them. Gaussian functions are fit with fixed widths of 10 \kms (blue) and 20 \kms (orange), to match the typical width retrieved from real detections. These distributions serve to illustrate the detection significance and sensitivities for each of the tested species, given the real noise properties of the cross-correlation functions. The blue distributions (10\kms)are wider than the orange distributions (20 \kms) because spurious fluctuations that are wider are less likely to occur at random. Red lines indicate the line-strengths of the detected species.}
	\label{fig:all_ccfs}
\end{figure*}

\noindent The time-averaged, one-dimensional CCF is then cropped into 15 smaller bins, within each of which a Gaussian profile is fit at a random location and with a fixed width of either 10 or 20 \kms. For each realization of the 1D CCF, this therefore yields 15 randomised measurements, and the experiment is repeated 5,000 times to yield 75,000 realizations of random Gaussian fits to the CCF of each species. The resulting distributions are plotted in Fig. \ref{fig:all_ccfs}. These figures show that the detected line strengths are significantly stronger than expected from random fluctuations occuring in the CCFs at radial velocities far away from the rest-frame of the system.

\clearpage

\section{Cross-correlation functions and bootstrap results}\label{app:results}
\begin{figure*}
    \includegraphics[width=0.9\textwidth]{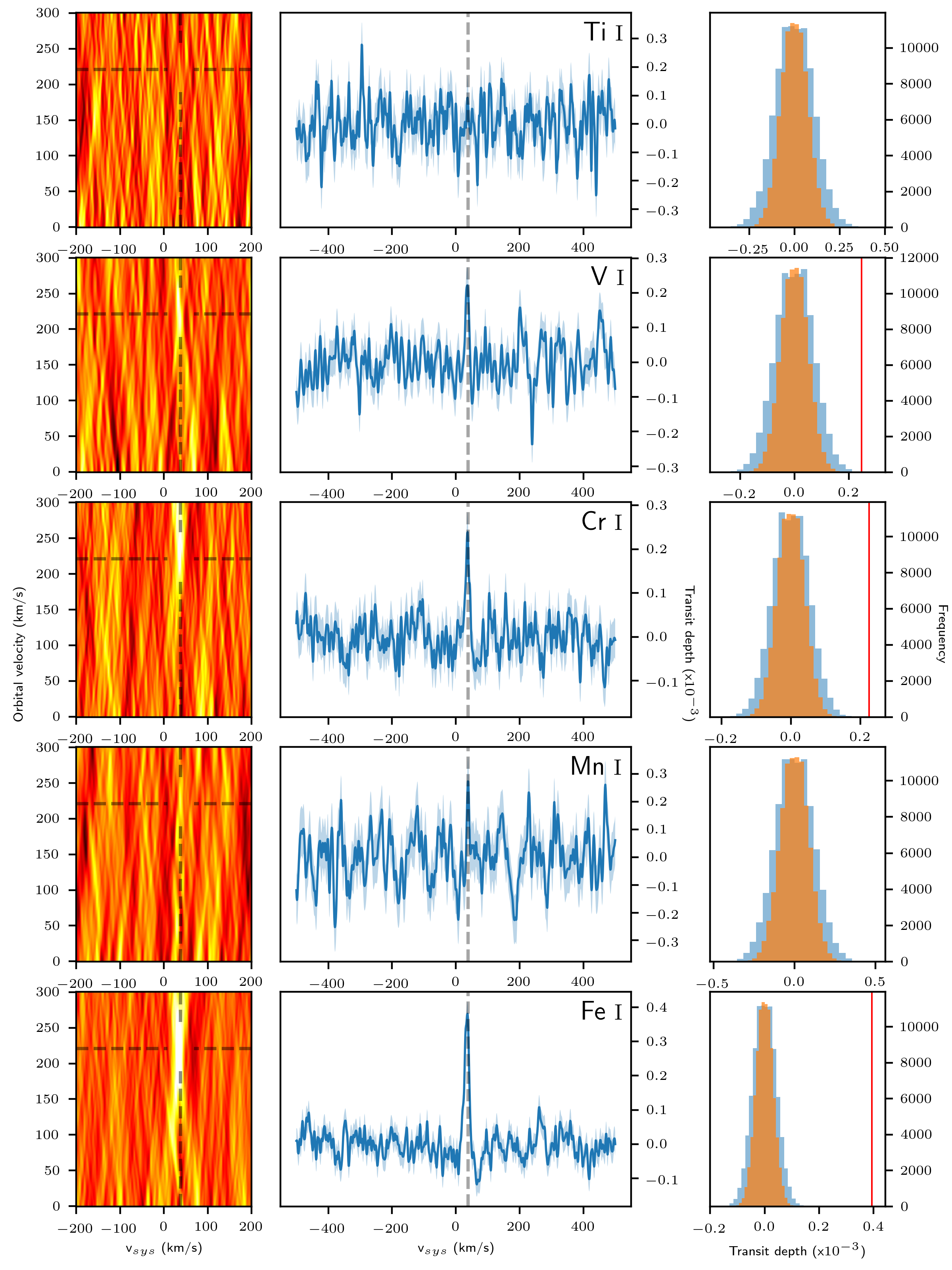}
\end{figure*}
\begin{figure*}
    \includegraphics[width=0.9\textwidth]{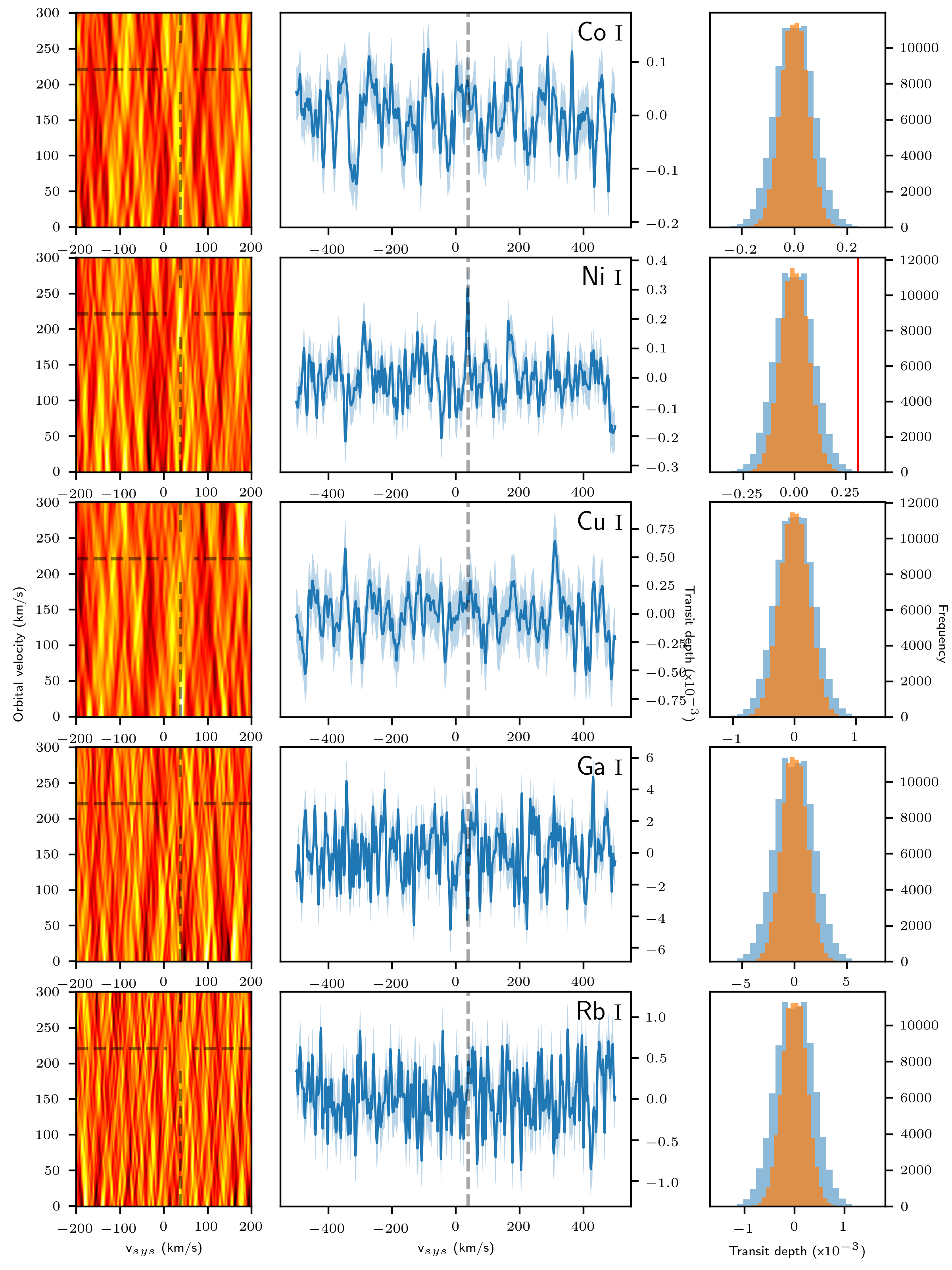}
\end{figure*}
\begin{figure*}
    \includegraphics[width=0.9\textwidth]{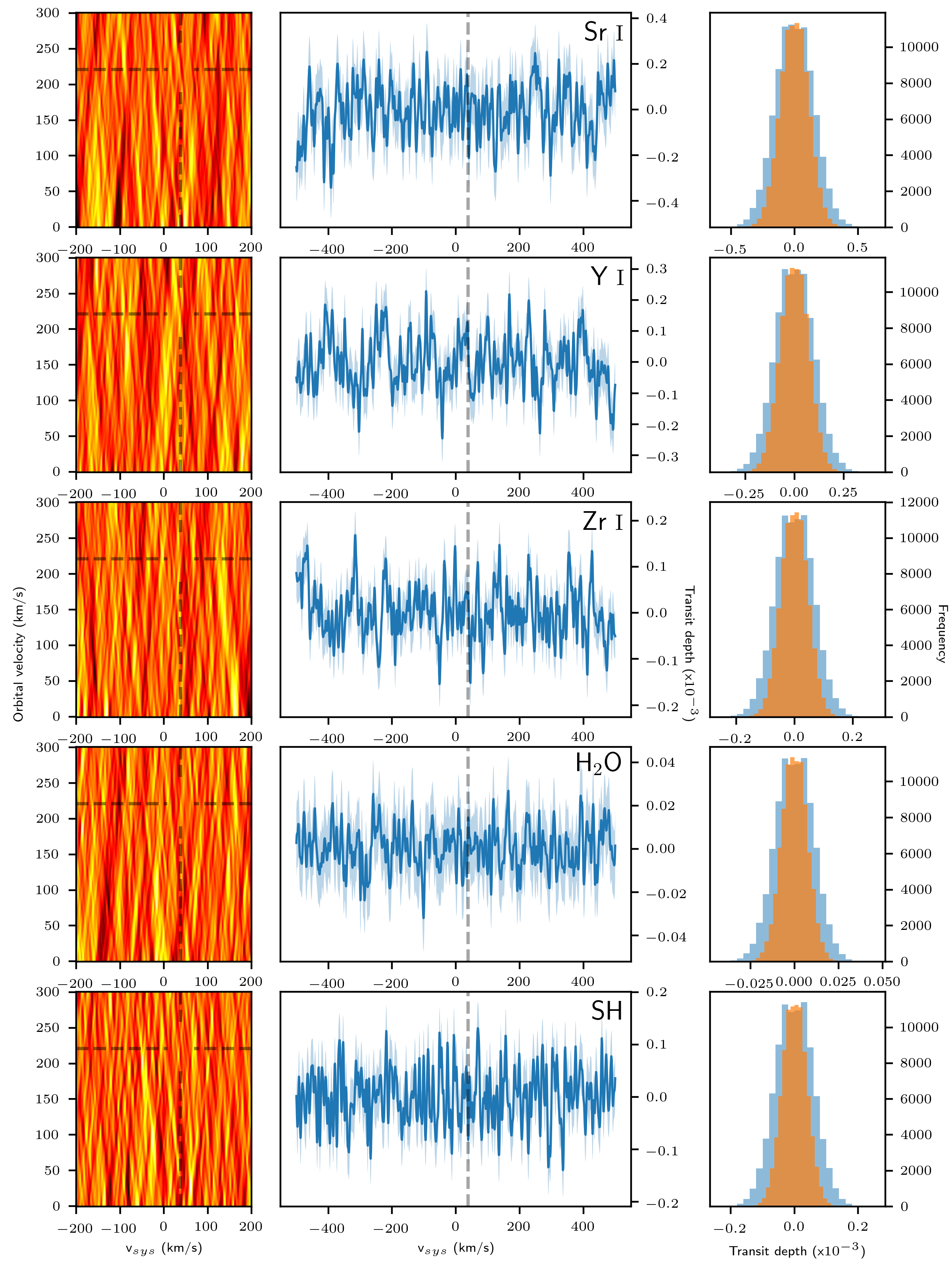}
\end{figure*}
\begin{figure*}
    \includegraphics[width=0.9\textwidth]{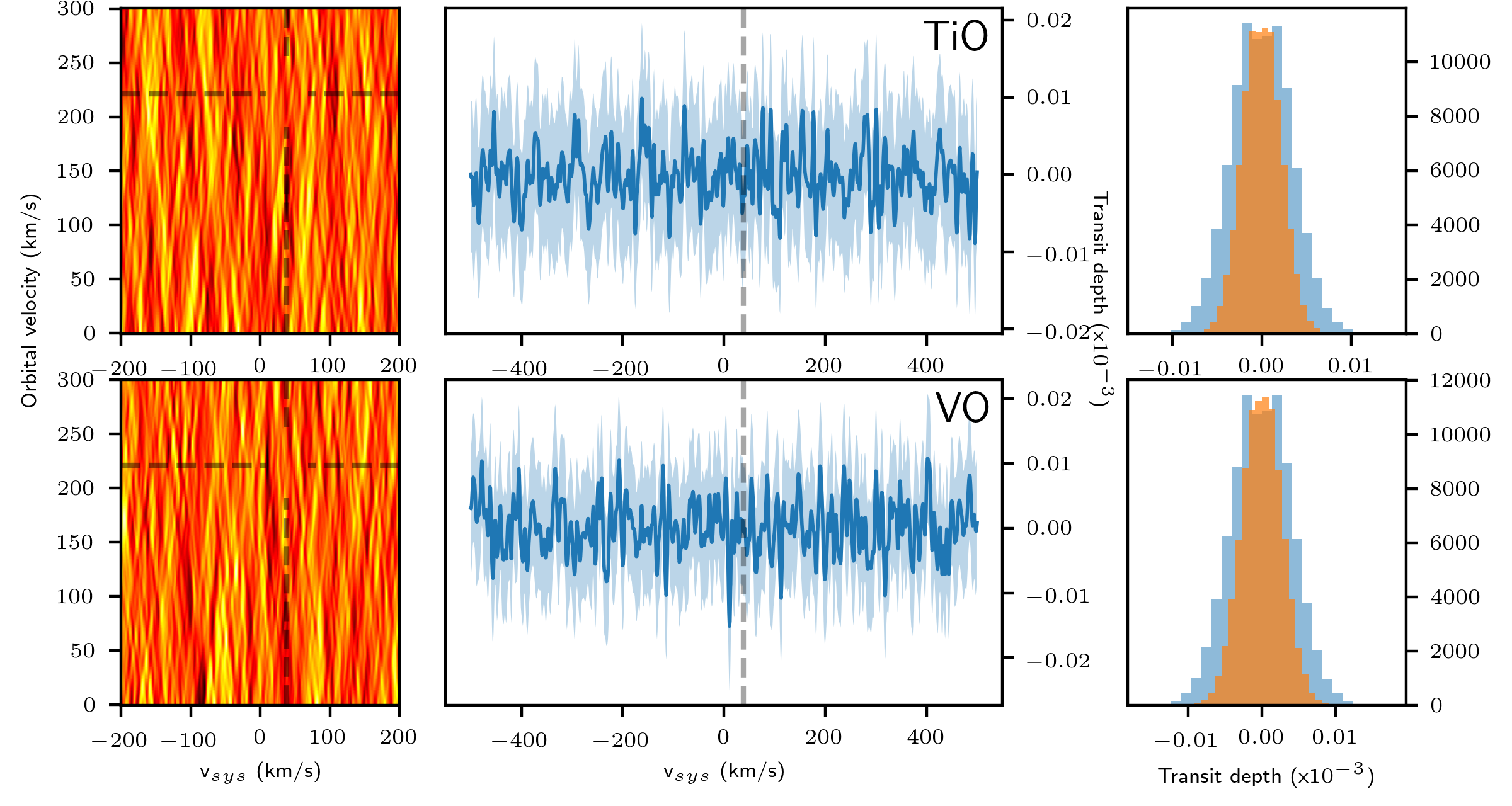}
\end{figure*}

\end{appendix}
\end{document}